\documentclass[superscriptaddress,amsmath,amssymb,prd,preprintnumbers,showpacs,twocolumn,nofootinbib]{revtex4-2}

\usepackage[colorlinks=true,pdfstartview=FitV,linkcolor=blue,citecolor=blue,urlcolor=blue,breaklinks=true]{hyperref}
\usepackage{graphicx}
\usepackage[T1]{fontenc} 
\usepackage{amssymb,amsmath,bm,natbib}
\usepackage{color}
\usepackage{slashed}
\usepackage{graphics}
\usepackage{graphicx}
\usepackage[utf8]{inputenc}
\usepackage[caption=false]{subfig}
\usepackage{hyperref}
\usepackage{url}
\usepackage{dsfont}
\usepackage{float}
\usepackage{cancel}
\usepackage{units}
\usepackage{blindtext}
\usepackage[utf8]{inputenc}
\usepackage{upgreek}
\usepackage{booktabs}
\usepackage[dvipsnames,table,xcdraw]{xcolor}
\usepackage{enumerate}
\usepackage{mathtools}
\usepackage{soul,color}
\usepackage[normalem]{ulem}

\newcommand{\orcid}[1]{\href{https://orcid.org/#1}{\includegraphics[width=10pt]{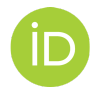}}}

\begin{document}

\title{Alternative classical Lagrangians for the Standard-Model Extension}

\author{Jo\~{a}o A.A.S.\ Reis\orcid{0000-0002-2831-5317}}
\email{joao.reis@uesb.edu.br}
\affiliation{Departamento de Ci\^{e}ncias Exatas e Naturais, \\ Universidade Estadual do Sudoeste da Bahia, Itapetinga (BA), 45700-000, Brazil}

\author{Marco Schreck\orcid{0000-0001-6585-4144}}
\email{marco.schreck@ufma.br}
\affiliation{Departamento de F\'{i}sica, Universidade Federal do Maranh\~{a}o, Campus Universit\'{a}rio do Bacanga, S\~{a}o Lu\'{i}s (MA), 65085-580, Brazil}

\author{Ronaldo Thibes\orcid{0000-0002-4291-8034}}
\email{thibes@uesb.edu.br}
\affiliation{Departamento de Ci\^{e}ncias Exatas e Naturais, \\ Universidade Estadual do Sudoeste da Bahia, Itapetinga (BA), 45700-000, Brazil}

\begin{abstract}

The current paper introduces classical, relativistic Lagrangians for point-particle analogs to the field theory description of the Standard-Model Extension (SME) for Lorentz violation. Lagrangians of a form alternative to those derived and studied in previous works are in the spotlight. Interestingly, they have well-defined massless limits, which makes them suitable for describing classical-particle analogs of photons subject to Lorentz violation. We first deal with different types of Dirac fermion coefficients, followed by various configurations of the SME photon sector. The Lagrangians are accompanied by constraints that we treat properly using the techniques due to Dirac. The results encountered may find application in photon propagation through gravitational fields in the presence of spacetime symmetry violation. Connections to Finsler geometry are likely to exist.

\end{abstract}

\pacs{11.30.Cp, 03.30.+p, 45.50.-j}
\keywords{Lorentz violation; Special relativity, Particle kinematics}
\maketitle

\section{Introduction}

Violations of the fundamental spacetime symmetries may be a signal for physics at the Planck scale, where quantum effects and the gravitational interaction are expected to be equally significant. In fact, theoretical evidence for such violations has emerged in strings~\cite{Kostelecky:1988zi,Kostelecky:1989jp,Kostelecky:1989jw,Kostelecky:1991ak,Kostelecky:1994rn,Kostelecky:1995qk,Mavromatos:2007xe,Mavromatos:2002re} loop quantum gravity \cite{Gambini:1998it,Alfaro:1999wd,Alfaro:2001rb,Alfaro:2002xz,Bojowald:2004bb}, spacetime foam models~\cite{Wheeler:1957mu,Hawking:1978pog,Klinkhamer:2003ec,Bernadotte:2006ya,Hossenfelder:2014hha,Li:2023wlo}, approaches to a noncommutative spacetime structure \cite{Amelino-Camelia:1999jfz,Carroll:2001ws}, chiral field theories on topologically nontrivial spacetime manifolds~\cite{Klinkhamer:1998fa,Klinkhamer:1999zh,Klinkhamer:2002mj,Ghosh:2017iat}, and alternative ultraviolet completions of general relativity such as Ho\v{r}ava-Lifshitz gravity~\cite{Horava:2009uw}.

The Standard-Model Extension (SME) \cite{Colladay:1996iz,Colladay:1998fq,Kostelecky:2000mm} was conceived as an effective field theory framework for experimental tests of Lorentz violation. As a classical field theory, it is capable of describing propagating plane waves and wave packets through superposition. After all, field theory was originally introduced to model continuous mechanical systems with an infinite number of degrees of freedom. What the SME cannot do is parametrize Lorentz violation for a classical, relativistic, pointlike particle. The closest a field theory comes to describing the latter is by constructing a wave packet of ever decreasing width, whose limit is the Dirac delta function in position space. However, doing so comes with several impasses. First, a Dirac function in position space involves an infinite number of plane waves with different frequencies. Second, for dispersion relations that are nonlinear in the wave number, the width of a wave packet increases over time.

There are several reasons why a parametrization of Lorentz violation for classical, pointlike particles can be valuable. The first is to be found in phenomenology. Some experimental tests of the gravitational laws, especially tests of the universality of free fall~\cite{Nobili:2013,Sanders:2000,Overduin:2012uk,Nobili:2012uj,MICROSCOPE:2022doy}, are carried out with macroscopic test masses; see also Refs.~\cite{Kostelecky:2010ze,Tasson:2017qsd}. A field theory approach is far from suitable for handling the physical behavior of such a system, but only classical mechanics allows for doing so. The second reason is a mathematical one. Classical point-particle Lagrangians are, in principle, path length functionals, whereupon they establish a direct connection to pseudo-Riemannian geometry. Modified path length functionals give rise to Finsler structures, which describe extensions of Riemannian geometry~\cite{Riemann:1868,Finsler:1918,Cartan:1933,Matsumoto:1986,Chern:1996,Bao:2000,Bao:2004}. The latter can have appealing properties that are interesting within pure mathematics as well as in applications~\cite{Antonelli:1993,Kozma:2003,Bucataru:2007}. In fact, Finsler geometry and modified-gravity models resting upon it experienced a \textit{renaissance} some years ago. The reader may consult the extensive reviews~\cite{Bubuianu:2018wry,Vacaru:2018yha,Bubuianu:2025zjb} providing detailed information on the history, development, mathematical properties, and applications to gravity.

To construct modified point-particle settings in the presence of Lorentz violation, one could introduce new background fields at the level of the Lagrangian for a pointlike particle. However, these background fields would have no obvious relationship to those of the SME, whose physics has been investigated since the construction of the minimal SME in 1998~\cite{Colladay:1996iz,Colladay:1998fq,Kostelecky:2000mm}. A more suitable approach is to start from the SME and to develop a mapping procedure to classical Lagrangians that now involve the same background fields already used in the SME. Such a procedure was successfully introduced for the minimal SME in 2010~\cite{Kostelecky:2010hs}. Afterwards, it found application in multiple coefficient sets of the SME fermion sector~\cite{Colladay:2012rv,Schreck:2014ama,Russell:2015gwa,Colladay:2017bon}, and the relation to Finsler geometry was established, too~\cite{Kostelecky:2011qz,AlanKostelecky:2012yjr,Colladay:2015wra}. Moreover, the technique also turned out to be fruitful in the nonminimal SME~\cite{Schreck:2014hga,Schreck:2015seb,Reis:2017ayl}. Finally, a perturbative approach to construct this map emerged in Ref.~\cite{Edwards:2018lsn}, initially used within a modified scalar field theory and later extended to Dirac fermions~\cite{Schreck:2019mmr,Reis:2021ban}.

Some of the SME classical Lagrangians encountered before are known to occur in certain problems or have been subject to additional studies. For example, the $a_{\mu}$ and $b_{\mu}$-type structures play a crucial role in problems of classical mechanics and electrodynamics~\cite{Zermelo:1931,Foster:2015yta}. The authors of Refs.~\cite{Silva:2013xba,Silva:2019qzl} focus on particle kinematics based on yet another class of SME point-particle Lagrangians. Different field theories for scalar, vector, and Dirac fields based on Randers spacetime are defined and examined in Ref.~\cite{Silva:2020tqr}. In the latter paper, it is also shown how these settings reproduce several terms contained in the SME action.

Most of the investigations carried out in the previously mentioned papers are based on modifications of the relativistic Lagrangian $L_0=-m\sqrt{\dot{x}^2}$~\cite{Goldstein:1980, Henneaux:1992ig}. The latter is valid for a pointlike particle of mass $m$ propagating along the trajectory $x^{\mu}=x^{\mu}(\lambda)$ with four-velocity $\dot{x}^{\mu}$ in terms of the parameter $\lambda$. Lagrangians of this form have been used in the SME to describe point-particle propagation in the presence of SME background fields.
Despite the interesting features of these Lagrangians, they also have several disadvantages. These are already evident from the standard Lagrangian $L_0$ and are inherited by their Lorentz-violating counterparts. First, $L_0$ does not possess a useful massless limit, i.e., it is unsuitable for pointlike particles without mass. Second, $L_0$ is nondifferentiable on the nullcone $\dot{x}^2=0$ in position space, which again reveals serious issues in the massless regime. Third, the canonical momentum associated with $L_0$ cannot be inverted for the particle velocity, disclosing the presence of primary constraints related to the dynamical variables.

Interestingly, a point-particle Lagrangian alternative to $L_0$ was proposed decades ago~\cite{Brink:1976sz,Brink:1976uf,Sundermeyer:1982,Tong:2009np,Morales:2017rlk} and motivated the form of the action for certain string field theories. This Lagrangian does not exhibit the problematic features of $L_0$ mentioned before, but it has not found broad application within the SME. The objective of the current paper is to present this Lagrangian, outline its properties, and apply it to various sectors of the minimal SME. Some of the latter have not yet been treated satisfactorily to this point.

Different definitions of point-particle Lagrangians and pseudo-Finsler structures have been introduced over the years. Initial treatments of Finsler structures with Lorentzian signature date back to Asanov~\cite{Asanov:1985} and Beem~\cite{Beem:1970}. Asanov defines the Lagrangian as the square of a nonnegative Finsler structure, which is positively homogeneous of degree 1 in the velocity. He then reduces the tangent bundle of the Finsler manifold to those vectors that he calls admissible. This approach is only capable of describing timelike geodesics. In contrast, Beem defines the Lagrangian independently of any Finsler structure as a real and sufficiently smooth function that is positively homogeneous of degree 2 in the velocity. The latter approach allows for timelike, lightlike, and spacelike geodesics, making it also adequate for describing the propagation of light rays.

More recent treatments by L\"{a}mmerzahl and collaborators~\cite{Lammerzahl:2008di,Lammerzahl:2012kw} are based on Beem's definition, but they relax the smoothness requirement from the original paper. Finslerian extensions of the Klein-Gordon, Maxwell, and Dirac equations are proposed in Refs.~\cite{Itin:2014uia,Lammerzahl:2018lhw}. The authors of Refs.~\cite{Perlick:2000,Perlick:2005hz,Hasse:2019zqi} demonstrate how the propagation of light rays in scenarios involving material media and modified vacua can be described in a pseudo-Finsler setting.

Pfeifer and Wohlfarth extend Beem's original definition by permitting Lagrangians that are homogeneous in the velocity of degree $\geq 2$~\cite{Pfeifer:2011tk,Pfeifer:2011xi}. Moreover, they require the Lagrangian to satisfy a kind of reversibility condition. Based on the latter definition, Ref.~\cite{Pfeifer:2011ve} defines a scalar Finslerian field theory, and Finslerian extensions of Einstein's gravity theory are examined in Refs.~\cite{Pfeifer:2011xi,Hohmann:2018rpp}. Last but not least, S\'{a}nchez and collaborators use their own definition of a pseudo-Finsler structure~\cite{Javaloyes:2018lex}, which they compare to many of the existing approaches in the contemporary literature. They can look back on a significant amount of research on questions of causality in Finsler spacetimes and applications in special and general relativity \cite{Caponio:2009er,Javaloyes:2013ika,Caponio:2014gra,Javaloyes:2022hph,Sanchez:2025ifx}. The latter outcomes are also a driving force behind research in black-hole physics \cite{Li:2024rvy,Dehkordi:2025mhz} based on techniques of Finsler geometry.

Our approach to constructing point-particle Lagrangians with modified kinematics differs from the other procedures outlined previously. Our Lagrangians involve an auxiliary quantity known as the \textit{einbein}, which gives rise to a different constraint structure and has not been considered in any of the previous techniques. The Lagrangians to be found are not positively homogeneous in the velocity, but they are positively homogeneous at first degree when taking the \textit{einbein} into account. As shall become clear below, this construction works for massive particles and is more than suitable for describing massless-particle kinematics, too.

The paper is organized as follows. Section~\ref{sec:lorentz-invariant-sector} reviews the properties of the two classical Lagrangians that are known for a Lorentz-invariant setting. Section~\ref{sec:lorentz-violating-sector} is dedicated to developing extensions of the alternative Lagrangian for complementary coefficient sets of the SME fermion sector. In particular, we treat the $a_{\mu}$, $e_{\mu}$, $b_{\mu}$, and $H_{\mu\nu}$ coefficients. In Sec.~\ref{sec:constraint-analysis}, we delve into a detailed analysis of the constraints that accompany the previous Lagrangians. Section~\ref{eq:massless-lagrangians} shows how to construct classical analogs to photons based on the Lagrangians focused on to this point. Finally, we conclude in Sec.~\ref{eq:conclusions} and provide an outlook on intriguing problems that could be tackled within the approach presented. Appendices~\ref{app:reparametrization-invariance}, \ref{app:additional-lagrangians}, and \ref{app:quartic-em-dispersion} present worthwhile calculational details and additional results. Natural units with $\hbar=c=1$ are used unless otherwise stated. The Minkowski metric has signature $(+,-,-,-)$. Lorentz and spatial indices are denoted by Greek and Latin letters, respectively.

\section{Lorentz-invariant case}
\label{sec:lorentz-invariant-sector}

Our starting point is the point-particle descriptions governed by each of the following classical Lagrangians~\cite{Brink:1976sz,Brink:1976uf,Sundermeyer:1982,Morales:2017rlk,Henneaux:1992ig,Tong:2009np}:
\begin{subequations}
\begin{align}
\label{eq:lagrangian-standard-form-1}
L_0=L_0(x,\dot{x})&=-m\sqrt{\dot{x}^2}=-m\sqrt{g_{\mu\nu}(x)\dot{x}^{\mu}\dot{x}^{\nu}}\,, \\[1ex]
\label{eq:lagrangian-standard-form-2}
\tilde{L}_0=\tilde{L}_0(x,\dot{x})&=-\frac{1}{2}\left(\frac{\dot{x}^2}{\mathfrak{e}}+\mathfrak{e}m^2\right) \notag \\
&=-\frac{1}{2}\left(\frac{1}{\mathfrak{e}}g_{\varrho\sigma}(x)\dot{x}^{\varrho}\dot{x}^{\sigma}+\mathfrak{e}m^2\right)\,,
\end{align}
\end{subequations}
where $g_{\mu\nu}(x)$ is a generic spacetime metric and $\mathfrak{e}$ is a scalar density given on the particle trajectory and interpreted as an \textit{einbein}. Note that Eq.~\eqref{eq:lagrangian-standard-form-1} is positively homogeneous of degree 1 in the velocity, i.e., $L_0(x,\zeta\dot{x})=\zeta L_0(x,\dot{x})$ for $\zeta>0$. The crucial implication of this property is that the action
\begin{equation}
S_0=\int\mathrm{d}\lambda\,L_0(x,\dot{x})\,,
\end{equation}
remains invariant under reparametrizations of the particle trajectory. On the contrary, at first glance, Eq.~\eqref{eq:lagrangian-standard-form-2} does not share this property with $L_0$. However, one must take into account the presence of the \textit{einbein}. The latter transforms under reparametrizations in a suitable manner such that the action remains invariant, apart from a surface term; see Refs.~\cite{Brink:1976uf,Tong:2009np} and App.~\ref{app:reparametrization-invariance}. Consequently, $\tilde{L}_0$ is homogeneous of degree 1 as well. We will encounter further consequences of this important characteristic at a later point.

The Lagrangians $L_0$ and $\tilde{L}_0$ are known to describe the same classical kinematics, as recapitulated in the following. The Euler-Lagrange equations for $\tilde{L}_0$ require
\begin{subequations}
\begin{align}
\label{eq:canonical-momentum-L0tilde}
p_{\mu}:=-\frac{\partial \tilde{L}_0}{\partial\dot{x}^{\mu}}&=\frac{1}{\mathfrak{e}}g_{\mu\nu}(x)\dot{x}^{\nu}\,, \\[1ex]
\frac{\partial \tilde{L}_0}{\partial x^{\mu}}&=-\frac{1}{2\mathfrak{e}}\partial_{\mu}g_{\varrho\sigma}\dot{x}^{\varrho}\dot{x}^{\sigma}\,,
\end{align}
\end{subequations}
with the canonical momentum $p_{\mu}$. Note the minus sign in the definition of the latter, which is introduced to make the four-velocity and canonical momentum collinear in Minkowski spacetime for $g_{\mu\nu}=\eta_{\mu\nu}$. The Euler-Lagrange equations are then cast into the form:
\begin{subequations}
\label{eq:euler-lagrange-equations}
\begin{align}
-\frac{\mathrm{d}p_{\mu}}{\mathrm{d}\tau}&=-\dot{p}_{\mu}=\frac{\partial \tilde{L}_0}{\partial x}\,, \\[1ex]
\dot{p}_{\mu}&=\frac{1}{2\mathfrak{e}}\partial_{\mu}g_{\varrho\sigma}\dot{x}^{\varrho}\dot{x}^{\sigma}\,.
\end{align}
\end{subequations}
The \textit{einbein} is treated as an additional nondynamical degree of freedom. In particular,
\begin{subequations}
\begin{align}
\frac{\partial \tilde{L}_0}{\partial \dot{\mathfrak{e}}}&=0\,, \\[1ex]
\frac{\partial \tilde{L}_0}{\partial \mathfrak{e}}&=\frac{1}{2}\left(\frac{\dot{x}^2}{\mathfrak{e}^2}-m^2\right)\,,
\end{align}
\end{subequations}
which implies the constraint relation
\begin{equation}
\label{eq:constraint}
0\approx \dot{x}^2-(m\mathfrak{e})^2\,,\quad \mathfrak{e}^2\approx \frac{\dot{x}^2}{m^2}\,,\quad \mathfrak{e}\approx \frac{\sqrt{\dot{x}^2}}{m}\,,
\end{equation}
where `$\approx$' denotes ``weak equality'' in the sense of constraints~\cite{Henneaux:1992ig}. The latter allows us to eliminate the \textit{einbein} from the Euler-Lagrange equations~\eqref{eq:euler-lagrange-equations}, which provides the same results as for $L_0$. This procedure can also be applied directly to Eq.~\eqref{eq:lagrangian-standard-form-2}.  In that case, the Lagrangian $\tilde{L}_0$ subject to the constraint of Eq.~\eqref{eq:constraint} is reformulated as
\begin{align}
\tilde{L}_0|_{\text{con}}&=-\frac{1}{2}\left(\frac{\dot{x}^2}{\mathfrak{e}}+\mathfrak{e}m^2\right)=-\frac{1}{2\mathfrak{e}}\dot{x}^2-\frac{1}{2}\mathfrak{e}m^2 \notag \\
&=-\frac{1}{2}\frac{m}{\sqrt{x^2}}\dot{x}^2-\frac{1}{2}\frac{\sqrt{\dot{x}^2}}{m}m^2 \notag \\
&=-\frac{m}{2}\sqrt{\dot{x}^2}-\frac{m}{2}\sqrt{\dot{x}^2}=-m\sqrt{\dot{x}^2}=L_0\,,
\end{align}
which leads us to the point-particle Lagrangian of the first form, given by Eq.~\eqref{eq:lagrangian-standard-form-1}.

To develop the Hamiltonian formulation of the point-particle based on $\tilde{L}_0$, the constraint relation in Eq.~\eqref{eq:constraint} is, at first, ignored. Then, we can invert the canonical momentum of Eq.~\eqref{eq:canonical-momentum-L0tilde} for the particle velocity without problems:
\begin{equation}
\dot{x}^{\nu}=\mathfrak{e}g^{\nu\varrho}p_{\varrho}\,.
\end{equation}
A Legendre transformation implies the canonical Hamiltonian:
\begin{subequations}
\label{eq:hamiltonian-canonical-standard}
\begin{align}
\tilde{H}_0&=-p\cdot \dot{x}-\tilde{L}_0 \notag \\
&=-\frac{1}{\mathfrak{e}}g_{\mu\nu}\dot{x}^{\nu}\dot{x}^{\mu}+\frac{1}{2}\left(\frac{1}{\mathfrak{e}}g_{\varrho\sigma}\dot{x}^{\varrho}\dot{x}^{\sigma}+\mathfrak{e}m^2\right) \notag \\
&=\frac{1}{2}\left(-\frac{1}{\mathfrak{e}}g_{\mu\nu}\dot{x}^{\mu}\dot{x}^{\nu}+\mathfrak{e}m^2\right) \notag \\
&=\frac{1}{2}\left(-\mathfrak{e}g_{\mu\nu}g^{\mu\varrho}p_{\varrho}g^{\nu\sigma}p_{\sigma}+\mathfrak{e}m^2\right)=-\frac{\mathfrak{e}}{2}\mathcal{D}(p)\,,
\end{align}
with the left-hand side of the relativistic particle dispersion equation
\begin{equation}
\label{eq:dispersion-relativistic}
\mathcal{D}(p)=p^2-m^2\,.
\end{equation}
\end{subequations}
Interestingly, the Hamiltonian based on the Lagrangian of the second form, Eq.~\eqref{eq:lagrangian-standard-form-2}, is proportional to $\mathcal{D}(p)$, which is known to represent a primary constraint for the standard case of Lagrangian $L_0$ \cite{Sundermeyer:1982}.  In the present context, the equation $\mathcal{D}(p)\approx 0$ is, in fact, a secondary constraint for $\tilde{L}_0$ which follows naturally from the canonical treatment {\it \`a la Dirac}, representing Eq.~\eqref{eq:constraint} in phase space through the momentum relation of Eq.~\eqref{eq:canonical-momentum-L0tilde}. We will return to this point later.

The Hamilton equations based on the canonical Hamiltonian~\eqref{eq:hamiltonian-canonical-standard} are obtained as follows:
\begin{subequations}
\begin{align}
\dot{x}_{\mu}&=-\frac{\partial \tilde{H}_0}{\partial p_{\mu}}=\mathfrak{e}p^{\mu}\,, \\[1ex]
\dot{p}_{\mu}&=\frac{\partial \tilde{H}_0}{\partial x_{\mu}}=\frac{\mathfrak{e}}{2}\partial_{\mu}g_{\varrho\sigma}p^{\varrho}p^{\sigma}\,.
\end{align}
\end{subequations}
Inserting the first set into the second set of equations results in
\begin{equation}
\dot{p}_{\mu}=\frac{\mathfrak{e}}{2}\partial_{\mu}g_{\varrho\sigma}\frac{\dot{x}^{\varrho}}{\mathfrak{e}}\frac{\dot{x}^{\sigma}}{\mathfrak{e}}=\frac{1}{2\mathfrak{e}}\partial_{\mu}g_{\varrho\sigma}\dot{x}^{\varrho}\dot{x}^{\sigma}\,,
\end{equation}
which corresponds to the equations of motion~\eqref{eq:euler-lagrange-equations}, computed in the Lagrangian formulation. In the remainder of this article, we will refer to Eq.~\eqref{eq:lagrangian-standard-form-1} and modifications thereof as type-1 Lagrangians. Analogously, Eq.~\eqref{eq:lagrangian-standard-form-2} and its appropriate Lorentz-violating versions will be denoted as type-2 Lagrangians.

\subsection{Constraints}

As we have already explained briefly above, constraints play a pivotal role for relativistic classical-particle Lagrangians. Let us first recall the constraint properties for $L_0$ of Eq.~\eqref{eq:lagrangian-standard-form-1}. The latter has a single primary first-class constraint given by the relativistic dispersion equation $\mathcal{D}(p)\approx 0$, cf.~Eq.~\eqref{eq:dispersion-relativistic}. In fact, reparametrization invariance of the point-particle action is a gauge symmetry, which the presence of first-class constraints hints at~\cite{Henneaux:1992ig}.

Note that the canonical Hamiltonian associated with $L_0$ is identically equal to 0. The reason is that the canonical momentum is positively homogeneous of degree 0 in the velocity. Then,  $L\equiv -p\cdot\dot{x}$ holds as an identity, which is a critical equation for the mapping procedure between field theory wave packets and classical point-particle Lagrangians in the SME; see Ref.~\cite{Kostelecky:2010hs}.

An important benefit from a consistent constraint analysis following Dirac and Bergmann~\cite{Dirac:1950pj,Anderson:1951ta} is the determination of the number of physical degrees of freedom of a system. The latter is computed from the generic formula~\cite{Henneaux:1992ig}
\begin{equation}
\label{eq:number-degrees-freedom}
N_{\mathrm{dof}}=\frac{1}{2}(N_{\mathrm{ph}}-2\times N_1-N_2)\,,
\end{equation}
where $N_{\mathrm{ph}}$ is the number of phase space variables and $N_{1,2}$ are the numbers of first- and second-class constraints, respectively. Since there are 8 phase space variables $x^{\mu}$ and $p_{\mu}$ and a single first-class constraint, $N_{\mathrm{dof}}=3$. This is the number of translational degrees of freedom of a classical, pointlike particle, as expected.

Now, gauge invariance of $L_0$ is broken when choosing a particular parametrization, e.g., via proper time. The dispersion equation then becomes a second-class constraint and the gauge fixing condition is second-class, too~\cite{Henneaux:1992ig}. Hence, the previous first-class constraint is replaced by two second-class constraints, which does not change the number of physical degrees of freedom according to Eq.~\eqref{eq:number-degrees-freedom}.

Let us now turn to $\tilde{L}_0$ in Eq.~\eqref{eq:lagrangian-standard-form-2}, which also possesses a singular Hessian characterizing a constrained system according to Dirac and Bergmann. It is worthwhile to mention that the introduction of the \textit{einbein} $\mathfrak{e}$ as an auxiliary time-evolving variable for the type-2 Lagrangian leads to additional constraints in phase space, which should be carefully dealt with. In terms of Dirac's nomenclature, associated with Eq.~\eqref{eq:lagrangian-standard-form-2}, we have two first-class constraints in phase space given by
\begin{subequations}
\label{eq:constraints-L0tilde}
\begin{align}
\label{eq:constraint-1-L0tilde}
\phi_1&:=\mathfrak{p} \approx 0 \,, \\[1ex]
\label{eq:constraint-2-L0tilde}
\phi_2&:=\frac{\mathcal{D}(p)}{2} \approx 0  \,,
\end{align}
\end{subequations}
where $\mathfrak{p}$ denotes the canonical momentum conjugated to the \textit{einbein}. The first one, $\phi_1$, is a primary constraint signaling the lack of a fully independent dynamics for the \textit{einbein} whose derivative does not occur in Eq.~\eqref{eq:lagrangian-standard-form-2}. On the other hand, $\phi_2$ is a secondary constraint resulting from the Poisson bracket structure, namely,
\begin{equation}
\{ \phi_1, \tilde{H}_0 \} = \phi_2 \,,
\end{equation}
with the canonical Hamiltonian of Eq.~\eqref{eq:hamiltonian-canonical-standard}. The curly brackets denote the binary operation for the Poisson algebra in the extended phase space $(x^\mu,\mathfrak{e},p_\mu,\mathfrak{p})$. As long as we do not resort to Eq.~\eqref{eq:constraint} or, equivalently, Eq.~\eqref{eq:constraint-1-L0tilde}, the associated Hamiltonian $\tilde{H}_0$ of Eq.~\eqref{eq:hamiltonian-canonical-standard} is nontrivial. After all, $\tilde{L}_0$ is positively homogeneous of degree 2 in the velocity. This makes the canonical momentum positively homogeneous of degree 1, whereupon the identity $L\equiv -p\cdot\dot{x}$ that holds for $L_0$ loses its validity.

Since we introduced the \textit{einbein} as an auxiliary field into $\tilde{L}_0$ of Eq.~\eqref{eq:lagrangian-standard-form-2} and worked on an extended phase space including the momenta, one may wonder if the type-2 Lagrangian exhibits the correct number of degrees of freedom. We have found two first-class constraints~\eqref{eq:constraints-L0tilde} and an absence of second-class constraints. In that respect, considering a total of ten coordinates in phase space, the number of physical degrees of freedom, according to Eq.~\eqref{eq:number-degrees-freedom}, amounts to
\begin{equation}
N_{\mathrm{dof}}=\frac{1}{2}(10 - 2\times 2 - 0)=3\,,
\end{equation}
as we found for $L_0$. Therefore, Eq.~\eqref{eq:lagrangian-standard-form-1} is expected to describe the same dynamics as does Eq.~\eqref{eq:lagrangian-standard-form-2}.

The presence of such constraints in phase space alters the canonical structure of the model, whose proper dynamical evolution requires the knowledge and appropriate use of Dirac brackets. We shall discuss this aspect in more detail within the Lorentz-violating regime, our main point of interest, in the forthcoming sections.

\section{Lorentz-violating regime}
\label{sec:lorentz-violating-sector}

In what follows, we will propose suitable extensions of the Lagrangian~\eqref{eq:lagrangian-standard-form-2}, whose equivalence to the known Lagrangians of type 1 will be established. Initially, we keep the particle mass $m$ to be able to consult the massive Lagrangians known for fermion sector coefficients of the minimal SME. In the second part of the current section, the mass will be dropped, allowing for a description of classical point-particle analogs for Weyl fermions as well as photons later.

\subsection{Massive classical-particle analogs}

The following field-theory action describes the minimal-SME fermion sector \cite{Colladay:1996iz,Colladay:1998fq,Kostelecky:2000mm}:
\begin{subequations}
\begin{align}
S_D&=\int\mathrm{d}^4x\,\mathcal{L}_D\,, \displaybreak[0]\\[1ex]
\mathcal{L}&=\frac{1}{2}\overline{\psi}(\widehat{\Gamma}^{\nu}\mathrm{i}\partial_{\nu}-\widehat{M})\psi+\text{H.c.}\,, \displaybreak[0]\\[1ex]
\widehat{\Gamma}^{\nu}&=\gamma^{\nu}+c^{\mu\nu}\gamma_{\mu}+d^{\mu\nu}\gamma_5\gamma_{\mu} \notag \\
&\phantom{{}={}}+e^{\nu}\mathds{1}+\mathrm{i}f^{\nu}\gamma_5+\frac{1}{2}g^{\kappa\lambda\nu}\sigma_{\kappa\lambda}\,, \displaybreak[0]\\[1ex]
\widehat{M}&=m+a^{\mu}\gamma_{\mu}+b^{\mu}\gamma_5\gamma_{\mu}+\frac{1}{2}H^{\mu\nu}\sigma_{\mu\nu}\,,
\end{align}
\end{subequations}
where $\psi$ is a Dirac spinor field and $\overline{\psi}:=\psi^{\dagger}\gamma^0$ its Dirac conjugate for a spin-1/2 fermion of mass $m$. All fields are defined in Minkowski spacetime with metric tensor $\eta_{\mu\nu}$ of signature $(+,-,-,-)$. The Dirac matrices satisfy the Clifford algebra $\{\gamma^{\mu},\gamma^{\nu}\}=2\eta^{\mu\nu}\mathds{1}$, where $\mathds{1}$ is the identity matrix in spinor space. Moreover, we employ the chiral Dirac matrix $\gamma_5:=\mathrm{i}\gamma^0\gamma^1\gamma^2\gamma^3$ as well as the commutator $\sigma^{\mu\nu}:=\frac{\mathrm{i}}{2}[\gamma^{\mu},\gamma^{\nu}]$. The background fields $a_{\mu}$, $b_{\mu}$, and $H_{\mu\nu}$ parametrize Lorentz violation without the presence of derivatives, whereas $c_{\mu\nu}$, $d_{\mu\nu}$, $e_{\nu}$, $f_{\nu}$, and $g_{\kappa\lambda\nu}$ each comes with a derivative acting on the fields.

The cases to be studied first are the vector-valued SME coefficients $a_{\mu}$, $e_{\mu}$, and $b_{\mu}$ as well as the tensor-valued ones $H_{\mu\nu}$. These sectors are sufficiently different such that the analysis is not repetitive. We will propose modified classical Lagrangians of type 2 and show how they are related to the type-1 Lagrangians found in the contemporary literature~\cite{Kostelecky:2010hs,Kostelecky:2011qz,Colladay:2012rv,AlanKostelecky:2012yjr,Schreck:2014ama,Schreck:2014hga,Russell:2015gwa,Colladay:2015wra,Schreck:2015seb,Colladay:2017bon,Reis:2017ayl,Edwards:2018lsn,Schreck:2019mmr,Reis:2021ban}. Furthermore, the canonical Hamiltonians are computed, which reveals the relationship to the dispersion equations of the sectors considered.

\subsubsection{Coefficients $a_{\mu}$}

Let us start with the minimal $a_{\mu}$ coefficients. At the level of effective field theory, the single-particle $a_{\mu}$ coefficients are unphysical, since they can be absorbed into the Dirac field by a suitable coordinate transformation~\cite{Colladay:1996iz}. Nevertheless, they are still interesting for the following reasons. First, for either multiple particle species that couple to different background fields $a_{\mu}$ or in the presence of gravity, a subset of these coefficients is physical. Second, they pose the simplest extension of the Dirac sector that violates Lorentz invariance, making them a worthwhile starting point for studying Lorentz violation in fermions. Third, the classical type-1 Lagrangian derived for the $a_{\mu}$ coefficients is related to Randers space~\cite{Randers:1941gge}, which is the simplest extension of Riemannian geometry in a Finslerian setting.

We propose the following \textit{ansatz} for the type-2 Lagrangian of Eq.~\eqref{eq:lagrangian-standard-form-2}:
\begin{align}
\label{eq:lagrangian-second-form-a}
\tilde{L}_a&=-\frac{1}{2}\left[\frac{1}{\mathfrak{e}}(\dot{x}+\mathfrak{e}a)^2+\mathfrak{e}(m^2-a^2)\right] \notag \\
&=-\frac{1}{2}\left(\frac{\dot{x}^2}{\mathfrak{e}}+2a\cdot\dot{x}+\mathfrak{e}m^2\right)\,.
\end{align}
The four-velocity is suitably shifted by the dimensionless four-vector $\mathfrak{e}a^{\mu}$ and the mass is redefined. Rewriting this form as done in the second line gives rise to a contribution linear in the velocity. Now, the relationship between the canonical momentum and the velocity reads
\begin{equation}
\label{eq:canonical-momentum-a}
p_{\mu}=-\frac{\partial \tilde{L}_a}{\partial\dot{x}^{\mu}}=\frac{\dot{x}_{\mu}}{\mathfrak{e}}+a_{\mu}\,.
\end{equation}
From
\begin{equation}
\frac{\partial \tilde{L}_a}{\partial\mathfrak{e}}=\frac{1}{2}\left(\frac{\dot{x}^2}{\mathfrak{e}^2}-m^2\right)\,,
\end{equation}
we obtain the constraint relation
\begin{equation}
0\approx \dot{x}^2-\mathfrak{e}^2m^2\,,\quad \mathfrak{e}\approx \frac{\sqrt{\dot{x}^2}}{m}\,.
\end{equation}
Inserting the latter into the Lagrangian of Eq.~\eqref{eq:lagrangian-second-form-a} provides
\begin{align}
\label{eq:lagrangian-first-form-a}
\tilde{L}_a|_{\text{con}}&=-\frac{1}{2}\left(\dot{x}^2\frac{m}{\sqrt{\dot{x}^2}}+2a\cdot\dot{x}+m\sqrt{\dot{x}^2}\right) \notag \\
&=-m\sqrt{\dot{x}^2}-a\cdot\dot{x}=L_a\,,
\end{align}
which corresponds to the Randers-type Lagrangian established in the literature \cite{Kostelecky:2010hs}. Inverting Eq.~\eqref{eq:canonical-momentum-a} for the velocity,
\begin{equation}
\dot{x}_{\mu}=\mathfrak{e}p_{\mu}-a_{\mu}\,,
\end{equation}
leads to the canonical Hamiltonian:
\begin{subequations}
\begin{align}
\tilde{H}_a&=-p\cdot \dot{x}-\tilde{L}_a \notag \\
&=-\mathfrak{e}p^2+\mathfrak{e}a\cdot p+\frac{1}{2}\left[\mathfrak{e}\left(\frac{\dot{x}}{\mathfrak{e}}+a\right)^2+\mathfrak{e}(m^2-a^2)\right] \notag \\
&=-\mathfrak{e}p^2+\mathfrak{e}a\cdot p+\frac{1}{2}\mathfrak{e}p^2+\frac{\mathfrak{e}}{2}m^2-\frac{\mathfrak{e}}{2}a^2 \notag \\
&=-\frac{\mathfrak{e}}{2}\mathcal{D}_a(p)\,,
\end{align}
with the left-hand side of the SME dispersion equation for the $a_{\mu}$ coefficients \cite{Colladay:1996iz,Kostelecky:2000mm},
\begin{equation}
\label{eq:dispersion-a}
\mathcal{D}_a(p)=(p-a)^2-m^2\,.
\end{equation}
\end{subequations}
The latter canonical Hamiltonian is an appropriate extension of the standard result found in Eq.~\eqref{eq:hamiltonian-canonical-standard}. Therefore, we conclude that $\tilde{L}_a$ and $L_a$, the latter of which was originally obtained from the modified dispersion equation, describe the same dynamics. Note that the standard mass shell is shifted by $-a^{\mu}$, which is inverse to the shift in position space performed in Eq.~\eqref{eq:lagrangian-second-form-a}. This property indicates that the single-particle $a_{\mu}$ coefficients are, indeed, unphysical, since they can be removed from the dispersion equation by a suitable redefinition of the canonical momentum.

\subsubsection{Coefficients $e_{\mu}$}

The minimal $e_{\mu}$ coefficients are vector-like and cannot be removed by a field redefinition. In the context of extensions of the Standard Model, though, they are incompatible with the gauge structures $\mathit{SU}(2)_L$ and $\mathit{SU}(3)_c$. Nevertheless, they are interesting to consider in an extended quantum electrodynamics. The known type-1 Lagrangian for $e_{\mu}$ \cite{Kostelecky:2010hs} is of Randers type; cf.~Eq.~\eqref{eq:lagrangian-first-form-a}. However, it involves an additional effective metric $\Omega_{\mu\nu}$ that is different from the Minkowski metric and depends on $e_{\mu}$. Therefore, we propose the following type-2 Lagrangian:
\begin{subequations}
\label{eq:lagrangian-second-form-e}
\begin{align}
\tilde{L}_e&=-\frac{1}{2}\left(\frac{\Omega_{\mu\nu}\dot{x}^{\mu}\dot{x}^{\nu}}{\mathfrak{e}}-2\frac{me\cdot\dot{x}}{1-e^2}+\mathfrak{e}\frac{m^2}{1-e^2}\right) \notag \\
&=-\frac{1}{2}\bigg[\frac{1}{\mathfrak{e}}\left(\dot{x}-\mathfrak{e}\frac{me}{1-e^2}\right)^2+\frac{1}{\mathfrak{e}}\frac{(e\cdot\dot{x})^2}{1-e^2} \notag \\
&\phantom{{}={}}\hspace{0.7cm}+\mathfrak{e}\frac{(1-2e^2)m^2}{(1-e^2)^2}\bigg]\,,
\end{align}
with
\begin{equation}
\Omega_{\mu\nu}=\eta_{\mu\nu}+\frac{e_{\mu}e_{\nu}}{1-e^2}\,.
\end{equation}
\end{subequations}
The canonical momentum associated with $x^{\mu}$ is of the form
\begin{equation}
\label{eq:canonical-momentum-p}
p_{\mu}=-\frac{\partial \tilde{L}_e}{\partial\dot{x}^{\mu}}=\Omega_{\mu\nu}\frac{\dot{x}^{\nu}}{\mathfrak{e}}-\frac{me_{\mu}}{1-e^2}\,,
\end{equation}
and we also obtain that
\begin{equation}
\frac{\partial \tilde{L}_e}{\partial\mathfrak{e}}=\frac{1}{2}\left(\Omega_{\mu\nu}\frac{\dot{x}^{\mu}\dot{x}^{\nu}}{\mathfrak{e}^2}-\frac{m^2}{1-e^2}\right)\,.
\end{equation}
As before, the latter implies a constraint relation, which can be solved for the \textit{einbein}:
\begin{subequations}
\begin{align}
0&\approx \Omega_{\mu\nu}\dot{x}^{\mu}\dot{x}^{\nu}-\frac{\mathfrak{e}^2m^2}{1-e^2}\,, \\[1ex]
\mathfrak{e}&\approx\frac{\sqrt{1-e^2}}{m}\sqrt{\Omega_{\mu\nu}\dot{x}^{\mu}\dot{x}^{\nu}}\,.
\end{align}
\end{subequations}
Eliminating the \textit{einbein} from Eq.~\eqref{eq:lagrangian-second-form-e} provides
\begin{align}
\tilde{L}_e|_{\text{con}}&=-\frac{1}{2}\bigg[\Omega_{\mu\nu}\dot{x}^{\mu}\dot{x}^{\nu}\frac{m}{\sqrt{1-e^2}}\frac{1}{\sqrt{\Omega_{\mu\nu}\dot{x}^{\mu}\dot{x}^{\nu}}}-2\frac{me\cdot\dot{x}}{1-e^2} \notag \\
&\phantom{{}={}}\hspace{0.5cm} +\sqrt{\Omega_{\mu\nu}\dot{x}^{\mu}\dot{x}^{\nu}}\frac{\sqrt{1-e^2}}{m}\frac{m^2}{1-e^2}\bigg] \notag \\
&=-\frac{1}{2}\bigg[\frac{m}{\sqrt{1-e^2}}\sqrt{\Omega_{\mu\nu}\dot{x}^{\mu}\dot{x}^{\nu}}-2\frac{me\cdot\dot{x}}{1-e^2} \notag \\
&\phantom{{}={}}\hspace{0.5cm}+\frac{m}{\sqrt{1-e^2}}\sqrt{\Omega_{\mu\nu}\dot{x}^{\mu}\dot{x}^{\nu}}\bigg] \notag \\
&=-\frac{m}{\sqrt{1-e^2}}\sqrt{\Omega_{\mu\nu}\dot{x}^{\mu}\dot{x}^{\nu}}+\frac{me\cdot\dot{x}}{1-e^2}\,,
\end{align}
which reproduces the known type-1 Lagrangian for the $e_{\mu}$ coefficients~\cite{Kostelecky:2010hs}.

To compute the canonical Hamiltonian, the effective metric $\Omega_{\mu\nu}$ must be inverted. Interestingly, the inverse is quite simple:
\begin{equation}
\Omega^{\mu\nu}=\left(\eta_{\mu\nu}+\frac{e_{\mu}e_{\nu}}{1-e^2}\right)^{-1}=\eta^{\mu\nu}-e^{\mu}e^{\nu}\,,
\end{equation}
which satisfies $\Omega_{\mu\nu}\Omega^{\nu\varrho}=\delta_{\mu}^{\phantom{\mu}\varrho}$, as can be checked quickly.
Thus, inverting Eq.~\eqref{eq:canonical-momentum-p} for the four-velocity leads to
\begin{equation}
\dot{x}^{\nu}=\mathfrak{e}\Omega^{\mu\nu}\left(p_{\mu}+\frac{me_{\mu}}{1-e^2}\right)\,.
\end{equation}
To compute the canonical Hamiltonian, it is instructive to employ the type-2 Lagrangian stated in the first line of Eq.~\eqref{eq:lagrangian-second-form-e}. Expressing the velocity in terms of the momentum results in
\begin{subequations}
\begin{align}
\tilde{H}_e&=-p\cdot \dot{x}-\tilde{L}_e=-\mathfrak{e}\Omega^{\mu\nu}\left(p_{\mu}+\frac{me_{\mu}}{1-e^2}\right)p_{\nu} \notag \\
&\phantom{{}={}}+\frac{\mathfrak{e}}{2}\left[\Omega^{\mu\nu}\left(p_{\mu}p_{\nu}-\frac{m^2}{(1-e^2)^2}e_{\mu}e_{\nu}\right)+\frac{m^2}{1-e^2}\right] \notag \\
&=-\frac{\mathfrak{e}}{2}\left[\Omega^{\mu\nu}\left(p+\frac{me}{1-e^2}\right)_{\mu}\left(p+\frac{me}{1-e^2}\right)_{\nu}-\frac{m^2}{1-e^2}\right] \notag \\
&=-\frac{\mathfrak{e}}{2}\left[p^2+2me\cdot p-(e\cdot p)^2-m^2\right] \notag \\
&=-\frac{\mathfrak{e}}{2}\mathcal{D}_e(p)\,,
\end{align}
with
\begin{equation}
\label{eq:dispersion-e}
\mathcal{D}_e(p)=p^2-(m-e\cdot p)^2\,.
\end{equation}
\end{subequations}
As before, the result involves the left-hand side of the modified dispersion equation, but now for the $e_{\mu}$ coefficients \cite{Kostelecky:2000mm}. What they do is tilt and deform the mass shell.

\subsubsection{Coefficients $b_{\mu}$}

The vector-like $b_{\mu}$ coefficients exhibit properties beyond those of the $a_{\mu}$ and $e_{\mu}$ coefficients studied previously. They lift the spin degeneracy in the dispersion relation for Dirac fermions such that there are two distinct dispersion relations depending on the spin projection. The dispersion equation itself is quartic in the momentum components and, in general, it does not factor into quadratic dispersion equations, as in the Lorentz-invariant regime or for the $a_{\mu}$ and $e_{\mu}$ coefficients. Nevertheless, there is a trick to factorize the dispersion equation at least partially, which we will revisit below.

The $b_{\mu}$ coefficients have two distinct classical type-1 Lagrangians that are determined in Ref.~\cite{Kostelecky:2010hs}. Therefore, we must also propose two type-2 Lagrangians, and we make the following \textit{ansatz}:
\begin{equation}
\label{eq:lagrangian-second-form-b}
\tilde{L}_b^{(\pm)}=-\frac{1}{2}\left(\frac{\dot{x}^2}{\mathfrak{e}}\pm 2\sqrt{(b\cdot \dot{x})^2-b^2\dot{x}^2}+\mathfrak{e}m^2\right)\,.
\end{equation}
The latter is motivated by the fact that the type-1 Lagrangians decompose into sums of two contributions: the standard Lagrangian $L_0$ and modifications governed by the characteristic Gramian expression $(b\cdot \dot{x})^2-b^2\dot{x}^2$ contained in a square root.

Note that these Lagrangians bear some similarity to those considered in Ref.~\cite{Colladay:2018jic}, which also involve the \textit{einbein} as a Lagrange multiplier. However, the first terms of the Lagrangians in the latter paper manifestly depend on the particle mass, which is not the case here. Furthermore, the dispersion equation is directly added to the Lagrangians as a constraint. Thus, their procedure differs significantly from ours.

Due to the form of the proposed type-2 Lagrangians, the constraint relation for the \textit{einbein} remains unmodified:
\begin{equation}
\frac{\partial \tilde{L}_b^{(\pm)}}{\partial\mathfrak{e}}=\frac{1}{2}\left(\frac{\dot{x}^2}{\mathfrak{e}^2}-m^2\right)\,,\quad \mathfrak{e}\approx \frac{\sqrt{\dot{x}^2}}{m}\,.
\end{equation}
Eliminating the \textit{einbein} from Eq.~\eqref{eq:lagrangian-second-form-b} leads us to the known type-1 Lagrangians obtained in Ref.~\cite{Kostelecky:2010hs}:
\begin{align}
\label{eq:lagrangian-first-form-b}
\tilde{L}_b^{(\pm)}|_{\mathrm{con}}&=-\frac{1}{2}\left(\dot{x}^2\frac{m}{\sqrt{\dot{x}^2}}\pm 2\sqrt{(b\cdot\dot{x})^2-b^2\dot{x}^2}+\frac{\sqrt{\dot{x}^2}}{m}m^2\right) \notag \\
&=-m\sqrt{\dot{x}^2}\mp\sqrt{(b\cdot\dot{x})^2-b^2\dot{x}^2}=L_b^{(\mp)}\,.
\end{align}
Note the sign switch of the square-root expression when going from the type-1 to the type-2 Lagrangians. In the following, we will observe the great advantage of using Eq.~\eqref{eq:lagrangian-second-form-b} instead of Eq.~\eqref{eq:lagrangian-first-form-b}. After all, the canonical momenta of the latter cannot be inverted for the velocity. Interestingly, it is possible to do so for the canonical momenta
\begin{equation}
p_{\mu}=-\frac{\partial \tilde{L}_b^{(\pm)}}{\partial\dot{x}^{\mu}}=\frac{\dot{x}_{\mu}}{\mathfrak{e}}\pm\frac{(b\cdot\dot{x})b_{\mu}-b^2\dot{x}_{\mu}}{\sqrt{(b\cdot\dot{x})^2-b^2\dot{x}^2}}\,,
\end{equation}
associated with Eq.~\eqref{eq:lagrangian-second-form-b}. The velocity in terms of the momentum is then given by
\begin{equation}
\dot{x}^{\mu}=\mathfrak{e}\left(p^{\mu}\mp\frac{(b\cdot p)b^{\mu}-b^2p^{\mu}}{\sqrt{(b\cdot p)^2-b^2p^2}}\right)\,.
\end{equation}
Each of the two Lagrangians $\tilde{L}_b^{(\pm)}$ has their own canonical Hamiltonian:
\begin{subequations}
\label{eq:Hb}
\begin{align}
\tilde{H}_b^{(\pm)}&=-p\cdot \dot{x}-\tilde{L}_b^{(\pm)} \notag \\
&=-\frac{\dot{x}^2}{\mathfrak{e}}\mp\frac{(b\cdot\dot{x})^2-b^2\dot{x}^2}{\sqrt{(b\cdot\dot{x})^2-b^2\dot{x}^2}}+\frac{1}{2}\frac{\dot{x}^2}{\mathfrak{e}}\notag \\
&\phantom{{}={}}\pm\sqrt{(b\cdot\dot{x})^2-b^2\dot{x}^2}+\frac{\mathfrak{e}}{2}m^2 \notag \\
&=-\frac{1}{2}\frac{\dot{x}^2}{\mathfrak{e}}+\frac{\mathfrak{e}}{2}m^2 \notag \\
&=\frac{\mathfrak{e}}{2}\bigg[-p^2\pm 2\frac{(b\cdot p)^2-b^2p^2}{\sqrt{(b\cdot p)^2-b^2p^2}} \notag \\
&\phantom{{}={}}\hspace{0.4cm}-\frac{b^4p^2-2b^2(b\cdot p)^2+(b\cdot p)^2b^2}{(b\cdot p)^2-b^2p^2}+m^2\bigg] \notag \\
&=-\frac{\mathfrak{e}}{2}\mathcal{D}_b^{(\mp)}(p)\,,
\end{align}
with
\begin{equation}
\label{eq:dispersions-b}
\mathcal{D}_b^{(\pm)}(p)=p^2-b^2-m^2\pm 2\sqrt{(b\cdot p)^2-b^2p^2}\,.
\end{equation}
\end{subequations}
The latter expressions, when multiplied with each other, provide the left-hand side of the complete dispersion equation for the $b_{\mu}$ coefficients \cite{Colladay:1996iz,Kostelecky:2000mm}:
\begin{equation}
\label{eq:quartic-dispersion-b}
\mathcal{D}_b^{(+)}(p)\mathcal{D}_b^{(-)}(p)=(p^2-b^2-m^2)^2-4(b\cdot p)^2+4b^2p^2\,,
\end{equation}
which is quartic.

\subsubsection{Coefficients $H_{\mu\nu}$}

The $H_{\mu\nu}$ coefficients are components of an antisymmetric $(4\times 4)$ matrix, i.e., there are six independent coefficients. It is reasonable to decompose them into two groups characterized by the following Lorentz scalars:
\begin{equation}
\label{definition-X-Y}
X:=\frac{1}{4}H^2\,,\quad Y:=\frac{1}{4}\tilde{H}\cdot H\,,
\end{equation}
with the dual $\tilde{H}_{\mu\nu}:=\frac{1}{2}\varepsilon_{\mu\nu\varrho\sigma}H^{\varrho\sigma}$, expressed via the four-dimensional Levi-Civita symbol $\varepsilon_{\mu\nu\varrho\sigma}$. For configurations with $Y=0$, the classical Lagrangians of type-1 were obtained in the literature. Similarly to the Lagrangians for the $b_{\mu}$ coefficients, the latter decompose into sums of the standard term and Gramian modifications~\cite{Kostelecky:2010hs}. Thus, we propose the following \textit{ansatz} for type-2 Lagrangians of the $H_{\mu\nu}$ coefficients with $Y=0$:
\begin{equation}
\label{eq:lagrangian-second-form-H}
\tilde{L}_H^{(\pm)}=-\frac{1}{2}\left[\frac{x^2}{\mathfrak{e}}\pm 2\sqrt{\dot{x}\cdot H\cdot H\cdot \dot{x}+2X\dot{x}^2}+\mathfrak{e}m^2\right]\,.
\end{equation}
The canonical momenta are
\begin{equation}
p_{\mu}=-\frac{\partial \tilde{L}_H^{(\pm)}}{\partial \dot{x}^{\mu}}=\frac{\dot{x}_{\mu}}{\mathfrak{e}}\pm \frac{(\dot{x}\cdot H\cdot H)_{\mu}+2X\dot{x}_{\mu}}{\sqrt{\dot{x}\cdot H\cdot H\cdot \dot{x}+2X\dot{x}^2}}\,.
\end{equation}
Similarly as for $b_{\mu}$, the \textit{einbein} remains unmodified, and we employ the standard constraint relation of Eq.~\eqref{eq:constraint} to eliminate it:
\begin{align}
\label{eq:lagrangian-first-form-H}
\tilde{L}_H^{(\pm)}|_{\mathrm{con}}&=-\frac{1}{2}\bigg(\dot{x}^2\frac{m}{\sqrt{\dot{x}^2}}\pm 2\sqrt{\dot{x}\cdot H\cdot H\cdot \dot{x}+2X\dot{x}^2} \notag \\
&\phantom{{}={}}\hspace{0.6cm}+\frac{\sqrt{\dot{x}^2}}{m}m^2\bigg) \notag \\
&=-m\sqrt{\dot{x}^2}\mp\sqrt{\dot{x}\cdot H\cdot H\cdot \dot{x}+2X\dot{x}^2}=L_H^{(\mp)}\,,
\end{align}
with the type-1 Lagrangians $L_H^{(\pm)}$ found in Ref.~\cite{Kostelecky:2010hs}. Again, it is possible to invert the momentum for the velocity:
\begin{equation}
\dot{x}^{\mu}=\mathfrak{e}\left(p^{\mu}\mp \frac{(p\cdot H\cdot H)^{\mu}+2Xp^{\mu}}{\sqrt{p\cdot H\cdot H\cdot p+2Xp^2}}\right)\,.
\end{equation}
The Legendre transformation of the modified type-2 Lagrangians provides
\begin{subequations}
\begin{align}
\tilde{H}_H^{(\pm)}&=-p\cdot \dot{x}-\tilde{L}_H^{(\pm)} \notag \displaybreak[0]\\
&=-\frac{\dot{x}^2}{\mathfrak{e}}\mp \frac{\dot{x}\cdot H\cdot H\cdot \dot{x}+2X\dot{x}^2}{\sqrt{\dot{x}\cdot H\cdot H\cdot \dot{x}+2X\dot{x}^2}}+\frac{1}{2}\frac{\dot{x}^2}{\mathfrak{e}} \notag \\
&\phantom{{}={}}\pm \sqrt{\dot{x}\cdot H\cdot H\cdot \dot{x}+2X\dot{x}^2}+\frac{\mathfrak{e}}{2}m^2 \notag \displaybreak[0]\\
&=-\frac{1}{2}\frac{\dot{x}^2}{\mathfrak{e}}+\frac{\mathfrak{e}}{2}m^2 \notag \displaybreak[0]\\
&=\frac{\mathfrak{e}}{2}\bigg[-p^2\pm 2\frac{p\cdot H\cdot H\cdot p+2Xp^2}{\sqrt{p\cdot H\cdot H\cdot p+2Xp^2}}+m^2 \notag \\
&\phantom{{}={}}-\frac{-2Xp\cdot H\cdot H\cdot p+4Xp\cdot H\cdot H\cdot p+4X^2p^2}{p\cdot H\cdot H\cdot p+2Xp^2}\bigg] \notag \\
&=-\frac{\mathfrak{e}}{2}\mathcal{D}_H^{(\pm)}(p)\,,
\end{align}
with the factors of the quartic dispersion equation
\begin{equation}
\label{eq:dispersions-H}
\mathcal{D}_H^{(\pm)}(p)=p^2+2X-m^2\pm 2\sqrt{p\cdot H\cdot H\cdot p+2Xp^2}\,.
\end{equation}
\end{subequations}
Here, we used that $(p\cdot H\cdot H)^2=-2X(p\cdot H\cdot H\cdot p)$, as long as $Y=0$.
The product of the latter two branches provides the left-hand side of the dispersion equation for the $H_{\mu\nu}$ coefficients with $Y=0$:
\begin{equation}
\mathcal{D}_H^{(+)}(p)\mathcal{D}_H^{(-)}(p)=(p^2-m^2+2X)^2-8Xp^2-4p\cdot H\cdot H\cdot p\,.
\end{equation}
A derivation of both type-1 and type-2 Lagrangians for $Y\neq 0$ has turned out to be challenging.

\subsection{Additional remarks}

With the findings for $a_{\mu}$, $e_{\mu}$, $b_{\mu}$, and $H_{\mu\nu}$ at hand, it has been shown that it is, in principle, possible to propose type-2 Lagrangians for modified Dirac fermions in the SME. Further cases can be studied, but they do not substantially differ from those already considered. Hence, they are relegated to App.~\ref{app:additional-lagrangians}. Here, we would like to make further comments on our results that may be of interest for the reader.

The procedure to compute classical-particle Lagrangians that was introduced in Ref.~\cite{Kostelecky:2010hs} does not work to derive modified type-2 Lagrangians based on the SME. This can be understood as follows. Since the canonical momentum for type-1 Lagrangians is positively homogeneous of degree 0 in the velocity, the dispersion equation arises as a primary constraint that the canonical momentum satisfies: $\mathcal{D}_{\text{LV}}(p)\approx 0$. Thus, $\mathcal{D}_{\text{LV}}(p)=0$ is one of the five equations that must be solved to obtain the modified classical type-1 Lagrangian for a particular configuration of SME coefficients; see Ref.~\cite{Kostelecky:2010hs}.

On the contrary, the canonical momentum of type-2 Lagrangians is not positively homogeneous of degree 0 in the velocity. Thus, $\mathcal{D}_{\text{LV}}(p)\approx 0$ does not arise as a primary constraint, but only as a secondary one because of the presence of the \textit{einbein}. Therefore, as we have seen, the canonical Hamiltonians for each of the cases studied are different from 0. In fact, we observed that
\begin{equation}
\tilde{L}_{\text{LV}}=-p\cdot\dot{x}-\frac{\mathfrak{e}}{2}\mathcal{D}_{\text{LV}}(p)\,,
\end{equation}
for spin-degenerate configurations and
\begin{equation}
\tilde{L}_{\text{LV}}^{(\pm)}=-p\cdot\dot{x}-\frac{\mathfrak{e}}{2}\mathcal{D}^{(\mp)}_{\text{LV}}(p)\,,
\end{equation}
for spin-nondegenerate ones with the quartic dispersion equation $\mathcal{D}_{\text{LV}}(p)=\mathcal{D}_{\text{LV}}^{(+)}(p)\mathcal{D}_{\text{LV}}^{(-)}(p)=0$. Thus, since the method of Ref.~\cite{Kostelecky:2010hs} relies on the dispersion equation being satisfied by the canonical momentum, it cannot give rise to the type-2 Lagrangians derived to this point.

Interestingly, as can be seen, the type-2 Lagrangians of Eqs.~\eqref{eq:lagrangian-second-form-a}, \eqref{eq:lagrangian-second-form-e}, \eqref{eq:lagrangian-second-form-b}, and \eqref{eq:lagrangian-second-form-H} possess well-defined massless limits. Therefore, $m\rightarrow 0$ implies classical-particle analogs to Weyl fermions modified by $a_{\mu}$, $e_{\mu}$, $b_{\mu}$ and $H_{\mu\nu}$-type background fields, respectively. Further examples of classical Lagrangians for massless particles subject to Lorentz violation will be discussed below.

\section{Dirac-Bergmann Constraint Analysis}
\label{sec:constraint-analysis}

In this section, we analyze the constraints by following Dirac and Bergmann for the proposed type-2 Lagrangians with coefficients $a_\mu$ and $b_\mu$. The corresponding computations for the Dirac brackets (DBs) are included.  Since we are proposing a set of brand-new parametrized relativistic particle Lagrangians consistent with Lorentz symmetry violation according to the SME classification, it is essential to understand their constrained dynamical evolution in phase space in full detail. This investigation also provides the symplectic structure of the DBs as a deformation of the usual Poisson bracket algebra.

The extended phase space enjoys a natural Poisson bracket algebra given by
\begin{subequations}
\label{43a}
\begin{align}
\{x^\mu, p_\nu \} &= \delta^\mu_{\phantom{\mu}\nu} \,, \\[1ex]
\{ \mathfrak{e}, \mathfrak{p} \} &=1 \,,
\end{align}
\end{subequations}
with the Poisson bracket of two quantities $A$ and $B$ defined as
\begin{equation}
\{A,B\}:=\sum_{i=1}^{N_{\mathrm{ph}}/2} \left(\frac{\partial A}{\partial Q^i}\frac{\partial B}{\partial P^i}-\frac{\partial A}{\partial P^i}\frac{\partial B}{\partial Q^i}\right)\,,
\end{equation}
where $Q^i$ and $P^i$ represent the $N_{\mathrm{ph}}$ canonical position and momentum variables, respectively. Here, $\{Q^i\}=\{x^{\mu},\mathfrak{e}\}$, $\{P^i\}=\{p_{\mu},\mathfrak{p}\}$, and $N_{\mathrm{ph}}=10$.

The latter algebra governs the Hamiltonian evolution of the system and guides the canonical quantization process in the case of unconstrained systems. For constrained systems, as in the current case, consistency demands the replacement of the above Poisson brackets by the corresponding DBs. They can be obtained from the constraint structure as follows.

\subsubsection{Coefficients $a_{\mu}$}

The Lagrangian given by Eq.~\eqref{eq:lagrangian-second-form-a} leads to two first-class constraints in phase space, namely,
\begin{subequations}
\begin{equation}
\phi_a^1:= \mathfrak{p}
\end{equation}
and
\begin{equation}
\phi_a^2 := \frac{\mathcal{D}_a(p)}{2}\,,
\end{equation}
\end{subequations}
with $\mathcal{D}_a(p)$ given by Eq.~\eqref{eq:dispersion-a}. To compute the DBs, we consider the two additional gauge-fixing conditions
\begin{subequations}
\label{eq:gauge-fixing-a}
\begin{equation}\label{x0}
\phi_a^3:=x^0
\end{equation}
and
\begin{equation}
\phi_a^4:= \mathfrak{e}{(p^0-a^0)}\,,
\end{equation}
\end{subequations}
which imply an invertible matrix of Poisson brackets among the constraints and gauge-fixing conditions:
\begin{equation}\label{PBMa}
\Delta^{ij}_a=\{ \phi_a^i, \phi_a^j \}\,,~~~i,j=1\dots 4\,.
\end{equation}
The entries for $\Delta^{ij}_a$ can be read off explicitly from Tab.~\ref{t1}, allowing the computation of the DBs. In fact, the DBs between two given phase space functions $F$ and $G$ are defined as
\begin{equation}
\label{DB}
\{F , G \}_{\mathrm{DB}}:= \{F , G \} - \{F,\phi_a^i \}
{\Delta^{-1}_a}_{ij} \{\phi_a^j , G \}\,.
\end{equation}
Here, ${\Delta^{-1}_a}_{ij}$ denotes the inverse of Eq.~\eqref{PBMa}, which can be written as
\begin{subequations}
\begin{equation}
\Delta_a^{-1}={(\det \Delta_a )}^{-1/2}\left(
\begin{array}{cc}
\begin{array}{cc}
0&-\mathfrak{e}\\
\mathfrak{e}&0
\end{array}&
\mathfrak{M}\\
-\mathfrak{M}^T
&
\begin{array}{cc}
0&0\\
0&0
\end{array}
\end{array}
\right)\,,
\end{equation}
in terms of the submatrix
\begin{equation}
\mathfrak{M} :=
\left(
\begin{array}{cc}
0&p^0-a^0\\p_0-a_0&0
\end{array}
\right)\,.
\end{equation}
\end{subequations}
Note further that
\begin{equation}
\det \Delta_a = (p^0-a^0)^4\,.
\end{equation}
Using Eq.~\eqref{DB}, the nonzero fundamental DBs in phase space are readily obtained as
\begin{subequations}\label{DBa}
\begin{align}
\{ x^\mu , p_\nu \}_{\mathrm{DB}} &= \delta^\mu_{\phantom{\mu}\nu} - \frac{p^\mu - a^\mu}{p^0-a^0}\delta^0_{\phantom{0}\nu}\,, \\[1ex]
\{ x^\mu , \mathfrak{e} \}_{\mathrm{DB}} &= \frac{\mathfrak{e}(p^\mu - a^\mu)}{(p^0-a^0)^2} - \frac{\mathfrak{e} \delta^\mu_{\phantom{\mu}0}}{{p^0-a^0}}\,.
\end{align}
\end{subequations}
The above DBs describe a dynamical evolution consistent with the constraint structure of the system.  In particular, it can be immediately seen from Eq.~\eqref{DBa} that we have $\{x^0,p_\mu\}_{\mathrm{DB}} \equiv 0 \equiv \{x^0,\mathfrak{e}\}_{\mathrm{DB}}$, in accordance with the gauge choice \eqref{x0}.  Similarly, all remaining constraints $\phi_a^i$ can now safely be taken as strongly equal to zero throughout all phase space.
\captionsetup{labelformat=empty}
\begin{table}[t]
\centering
{\small\begin{tabular}{l|cccc}
\toprule
&$\phi_a^1$&$\phi_a^2$&$\phi_a^3$&$\phi_a^4$\\
\midrule
$\phi_a^1$&$0$&$0$&$0$&$-p_0+a_0$\\[0.5ex]
$\phi_a^2$&$0$&$0$&$-p^0+a^0$&$0$\\[0.5ex]
$\phi_a^3$&$0$&$p^0-a^0$&$0$&$\mathfrak{e}$\\[0.5ex]
$\phi_a^4$&$p_0-a_0$&$0$&$-\mathfrak{e}$&$0$\\
\bottomrule
\end{tabular}}
\caption{Dirac constraint algebra for $a_{\mu}$.}
\label{t1}
\end{table}

\subsubsection{Coefficients $b_{\mu}$}

For the coefficients $b_{\mu}$, concerning the phase space evolution associated with the Hamiltonians~\eqref{eq:Hb}, we have
\begin{subequations}\label{eq:phib}
\begin{equation}\label{eq:phi1b}
\phi_{b}^1:= \mathfrak{p}
\end{equation}
as a primary constraint and
\begin{equation}\label{eq:phi2b}
\phi_{b\pm}^2:=\frac{\mathcal{D}^{(\mp)}(p)}{2}\,,
\end{equation}
\end{subequations}
with $\mathcal{D}^{(\pm)}(p)$ of Eq.~\eqref{eq:dispersions-b} as secondary ones.
The {stability condition} applied to Eq.~\eqref{eq:phi2b} does not lead to further constraints, and we have the relations
\begin{subequations}
\label{Abel}
\begin{align}
\{\phi_{b}^1, \tilde{H}_b^{(\pm)} \} &= \phi_{b\pm}^2 \,, \\[1ex]
\{\phi_{b\pm}^2, \tilde{H}_b^{(\pm)} \} &= 0 \,,\\[1ex]
\{\phi_{b}^1, \phi_{b\pm}^2 \} &= 0 \,,
\end{align}
\end{subequations}
which ensure the first-class nature of the constraints~\eqref{eq:phib}.
The fact that we have managed to find all natural constraints hidden in Eq. \eqref{eq:lagrangian-second-form-b} can be confirmed by writing the primary Hamiltonian,
\begin{equation}
\tilde{H}_b^{P(\pm)}=\tilde{H}_b^{(\pm)}+\xi {\phi_{b}^1}\,,
\end{equation}
in terms of a Lagrange multiplier $\xi$.  Then, due to Eq.~\eqref{Abel}, the stability condition for $\phi_{b\pm}^2$ leads to an identity, and $\xi$ remains undetermined.
\begin{table}[t]
\centering
{\small{\begin{tabular}{l|cccc}
\toprule
 & $\phi_b^1$ & $\phi_{b\pm}^2$ & $\phi_b^3$ & $\phi_{b\pm}^4$ \\
\midrule
$\phi_b^1$ & $0$ & $0$ & $0$ & $-N^{~0}_\pm$ \\[0.5ex]
$\phi_{b\pm}^2$ & $0$ & $0$ & $-N^{~0}_\pm$ & $0$ \\[0.5ex]
$\phi_b^3$ & $0$ & $N^{~0}_\pm$ & $0$ & $\mathfrak{e}M^{~0}_{\pm~0 }$ \\[0.5ex]
$\phi_{b\pm}^4$ & $N^{~0}_\pm$ & $0$ & $-\mathfrak{e}M^{~0}_{\pm~0 }$ & $0$ \\
\bottomrule
\end{tabular}}}
\caption{Dirac constraint algebra for $b_{\mu}$ in terms of the quantities defined in Eqs.~\eqref{eq:quantity-N} and \eqref{eq:quantity-M}.}
\label{t2}
\end{table}

Hence, to proceed with the calculation of the DBs, as in the previous case, we need to fix the gauge. A feasible gauge condition is given by
\begin{subequations}
\label{eq:gauge-fixing-b}
\begin{equation}
\phi_b^3:=x^0\,,
\end{equation}
and
\begin{equation}
\phi_{b\pm}^4:= \mathfrak{e}N^{~0}_\pm (b,p)\,,
\end{equation}
\end{subequations}
with
\begin{equation}
\label{eq:quantity-N}
N^{~\mu}_\pm (b,p) := p^\mu \mp \frac{(b\cdot p) {b^\mu} - b^2 p^\mu}{\sqrt{(b\cdot p)^2 - b^2 p^2}}\,.
\end{equation}
The latter furnish the nonzero Poisson brackets of the constraint algebra:
\begin{subequations}
\label{b-algebra}
    \begin{align}
        \{\phi_{b}^1 , \phi_{b\pm}^4\}&= - N_\pm^{~0} (b,p) 
        \,, \\[1ex]
        \{ \phi_{b\pm}^2, \phi_b^3\}&= - N_\pm^{~0} (b,p) 
        \,, \\[1ex]
        \{ \phi_b^3 , \phi_{b\pm}^4 \}&= \mathfrak{e} M^{~0}_{\pm~0} (b,p)\,,
    \end{align}
\end{subequations}
in terms of
\begin{align}
\label{eq:quantity-M}
M^{~\mu}_{\pm~\nu} &:= \delta^\mu_{\phantom{\mu}\nu}\mp \frac{\left[(b\cdot p)^2-b^2p^2\right]\left[{b^\mu b_\nu} - b^2 \delta^\mu_{\phantom{\mu}\nu}\right]}{\left[(b\cdot p)^2-b^2 p^2\right]^{3/2}}\nonumber\\
    &\phantom{{}={}}\pm \frac{\left[(b\cdot p) b_\nu - b^2 p_\nu\right]\left[(b\cdot p) b^\mu - b^2 p^\mu\right]}{\left[(b\cdot p)^2-b^2 p^2\right]^{3/2}}\,.
\end{align}
The entries of the matrices $\Delta_{b\pm}$ containing the Poisson brackets among the constraints are stated in Tab.~\ref{t2}, from which their inverses can be written as
\begin{equation}
\Delta_{b\pm}^{-1}=(N^{~0}_\pm)^{-2}
\left(
\begin{array}{cccc}
0&-\mathfrak{e}M^{~0}_{\pm~0 }&0& N^{~0}_\pm\\
\mathfrak{e}M^{~0}_{\pm~0 }&0& \ N^{~0}_\pm&0\\
0&- N^{~0}_\pm&0&0\\
-  N^{~0}_\pm&0&0&0
\end{array}
\right)\,.
\end{equation}
By plugging the above inverses into Eq.~\eqref{DB} evaluated for the current $b_{\mu}$ case, we obtain the nonzero fundamental DB relations
\begin{subequations}\label{DBb}
\begin{align}
\{ x^\mu , p_\nu \}_{\mathrm{DB}} &= \delta^\mu_{\phantom{\mu}\nu} -  \frac{N_\pm^{~\mu} }
{ N_\pm^{~0} }
 \delta^0_{\phantom{0}\nu}\,, \\[1ex]
\{ x^\mu , \mathfrak{e} \}_{\mathrm{DB}} &= \mathfrak{e} \frac{\left( N_\pm^{~\mu} M_{\pm~0}^{~0} - N_\pm^{~0} M_{\pm~0}^{~\mu} \right)}{(N_\pm^{~0})^2}\,,
\end{align}
\end{subequations}
which fully characterize the constrained dynamical evolution of the system in the chosen gauge. Here, we can see a similar structure, as compared to Eq.~\eqref{DBa} for the $a_{\mu}$ case. The \textit{einbein} has a nontrivial DB with the spatial part of the particle coordinates, and it can be checked that each constraint function or gauge-fixing condition in Eqs.~\eqref{eq:phib} and \eqref{eq:gauge-fixing-b} has vanishing DBs with any phase-space function, as it should.

Finally, the number of physical degrees of freedom for a relativistic pointlike particle subject to Lorentz violation follows from the general formula of Eq.~\eqref{eq:number-degrees-freedom}. According to the results of the previous constraint analysis, $N_{\mathrm{dof}}=3$ is established. The conclusion is that Lorentz-violating background fields in Minkowski spacetime do not alter the number of physical degrees of freedom of the particle. The situation is expected to change for particle propagation in curved spacetime manifolds. Such a study may provide additional insightful results, but shall be postponed to a future time.

The constraint analyses for $e_{\mu}$ and $H_{\mu\nu}$ parallel the previous ones, although they are technically more involved. Therefore, we refrain from going through them explicitly here.

\section{Massless classical-particle analogs}
\label{eq:massless-lagrangians}

The massless limits of the classical type-2 Lagrangians established previously demonstrate the major advantage of these Lagrangians in comparison to those of the SME literature~\cite{Kostelecky:2010hs,Kostelecky:2011qz,Colladay:2012rv,AlanKostelecky:2012yjr,Schreck:2014ama,Schreck:2014hga,Russell:2015gwa,Colladay:2015wra,Schreck:2015seb,Colladay:2017bon,Reis:2017ayl,Edwards:2018lsn,Schreck:2019mmr,Reis:2021ban}, which are based on the Lagrangian of Eq.~\eqref{eq:lagrangian-standard-form-1}. Unsurprisingly, modifications resting upon Eq.~\eqref{eq:lagrangian-standard-form-2} are even more valuable for pointlike particles that are massless from the start, such as classical photon analogs. In what follows, we will discuss how to construct such Lagrangians for specific coefficient subsets of the minimal-SME photon sector. The latter is described by the action~\cite{Colladay:1998fq,Kostelecky:2002hh,Bailey:2004na}:
\begin{subequations}
\label{eq:photon-sector-sme}
\begin{align}
S_{\upgamma}&=\int\mathrm{d}^4x\,\mathcal{L}_{\upgamma}\,, \\[1ex]
\mathcal{L}_{\upgamma}&=-\frac{1}{4}F_{\mu\nu}F^{\mu\nu}-\frac{1}{4}(k_F)^{\mu\nu\varrho\sigma}F_{\mu\nu}F_{\varrho\sigma} \notag \\
&\phantom{{}={}}+\frac{1}{2}(k_{AF})^{\mu}\varepsilon_{\mu\nu\varrho\sigma}A^{\nu}F^{\varrho\sigma}\,,
\end{align}
\end{subequations}
in terms of the electromagnetic field strength tensor $F_{\mu\nu}=\partial_{\mu}A_{\nu}-\partial_{\nu}A_{\mu}$ with the $\mathit{U}(1)$ gauge field $A_{\mu}$. Here, $(k_F)^{\mu\nu\varrho\sigma}$ is a tensorial background field of rank 4 that governs a CPT-even modification. The vector-valued background $(k_{AF})^{\mu}$ gives rise to a CPT-odd alteration.

In contrast to the Dirac fermion sector, classical point-particle descriptions for massless particles based on the SME were considered in a single article only, see Ref.~\cite{Schreck:2015dsa}. The latter paper adopts the eikonal formulation of classical electromagnetism to the modified electromagnetic sector of the SME. Here, we will extend these results based on Eq.~\eqref{eq:lagrangian-standard-form-2} without resorting to the eikonal approach.

In general, massless particles move along nullcones in position space, which may be modified by the SME. Since the first term of Eq.~\eqref{eq:lagrangian-standard-form-2} contains the left-hand side of the standard nullcone, we shall generalize the latter Lagrangian based on this observation.

\subsection{Degenerate quadratic dispersion}
\label{sec:single-nullcone}

First of all, we consider classical photon analogs that are spin-degenerate. Thus, each of the transverse polarization modes propagates along a single modified nullcone described by
\begin{equation}
\label{eq:modified-nullcone-single}
0=\Omega_{\mu\nu}\dot{x}^{\mu}\dot{x}^{\nu}\,,
\end{equation}
where $\Omega_{\mu\nu}=\text{const.}$ is an effective metric. We then propose a modified Lagrangian of the form:
\begin{equation}
\label{eq:massless-lagrangian-single-nullcone}
\tilde{L}_{\upgamma}=-\frac{1}{2\mathfrak{e}}\Omega_{\mu\nu}\dot{x}^{\mu}\dot{x}^{\nu}\,.
\end{equation}
The canonical momentum reads
\begin{equation}
p_{\mu}=-\frac{\partial L_{\upgamma}}{\partial\dot{x}^{\mu}}=\frac{1}{\mathfrak{e}}\Omega_{\mu\nu}\dot{x}^{\nu}\,.
\end{equation}
The latter can be inverted for the velocity to give
\begin{equation}
\dot{x}^{\nu}=\mathfrak{e}\Omega^{\nu\varrho}p_{\varrho}\,,
\end{equation}
where $\Omega^{\mu\nu}$ is the inverse effective metric, i.e., $\Omega^{\nu\varrho}\Omega_{\varrho\sigma}=\delta^{\nu}_{\phantom{\nu}\sigma}$. Moreover, the modified nullcone is incorporated as a constraint relation:
\begin{equation}
0\approx \Omega_{\mu\nu}\dot{x}^{\mu}\dot{x}^{\nu}\,.
\end{equation}
A Legendre transformation implies the canonical Hamiltonian:
\begin{subequations}
\begin{align}
\tilde{H}_{\upgamma}&=-p\cdot\dot{x}-\tilde{L}_{\upgamma}=-\frac{1}{2\mathfrak{e}}\Omega_{\mu\nu}\dot{x}^{\mu}\dot{x}^{\nu} \notag \\
&=-\frac{\mathfrak{e}}{2}\Omega_{\mu\nu}\Omega^{\mu\varrho}p_{\varrho}\Omega^{\nu\sigma}p_{\sigma}=-\frac{\mathfrak{e}}{2}\mathcal{D}_{\upgamma}(p)\,,
\end{align}
where
\begin{equation}
\label{eq:dispersion-kappa}
\mathcal{D}_{\upgamma}(p)=\Omega^{\varrho\sigma}p_{\varrho}p_{\sigma}\,,
\end{equation}
\end{subequations}
is the left-hand side of the modified quadratic dispersion equation for photons. This result shows that propagation in position space is governed by the effective metric $\Omega_{\mu\nu}$, whereas the dispersion equation is based on its inverse~$\Omega^{\mu\nu}$. If the nullcone has a smaller opening angle in position space as compared to the Lorentz-invariant case, massless particles propagate more slowly than the maximum attainable velocity of massive particles, unaffected by Lorentz violation. Then, the associated cone in energy-momentum space is wider, which implies that the energy at a fixed momentum is reduced compared to a Lorentz-invariant massless dispersion relation.

In the following, we will discuss different examples of SME settings that can be described by Eq.~\eqref{eq:massless-lagrangian-single-nullcone}. Since the modified dispersion relations have been investigated in much more detail than the corresponding modified nullcones in position space, we will be starting from the inverse effective metric $\Omega^{\mu\nu}$.

First of all, let us again look at modified fermions. Since these are now considered massless, we are, in principle, dealing with Weyl fermions, as already noted before. For example, if massless fermions are modified by the $c_{\mu\nu}$ coefficients, the dispersion equation is spin-degenerate and has the form $\mathcal{D}_{\upgamma}(p)=0$ with an inverse effective metric given by
\begin{subequations}
\begin{equation}
\Omega^{\mu\nu}=\eta^{\mu\nu}+2c^{(\mu\nu)}+c^{\mu\kappa}c_{\kappa}^{\phantom{\kappa}\nu}\,,
\end{equation}
with the symmetric part $c^{(\mu\nu)}$ of $c^{\mu\nu}$; see Ref.~\cite{Kostelecky:2000mm}. The latter is, at most, quadratic in the coefficients. The effective metric in position space at first order in the coefficients then reads
\begin{equation}
\Omega_{\mu\nu}=\eta_{\mu\nu}-2c_{(\mu\nu)}+\dots\,.
\end{equation}
\end{subequations}
The effect of the $c_{\mu\nu}$-type background field is to skew the Minkowski metric that governs massless-fermion kinematics in the standard case. Similarly, the effective metrics for a massless fermion affected by the minimal $f_{\mu}$ coefficients read
\begin{subequations}
\begin{align}
\Omega^{\mu\nu}&=\eta^{\mu\nu}-f^{\mu}f^{\nu}\,, \\[1ex]
\Omega_{\mu\nu}&=\eta_{\mu\nu}+f_{\mu}f_{\nu}+\dots\,,
\end{align}
\end{subequations}
in position and momentum space, respectively. Note that the latter relationships are already of second order in $f_{\mu}$, which is a peculiarity that holds for these coefficients~\cite{Altschul:2006ts}.

Moreover, dispersion equations $D_{\upgamma}(p)=0$ with $D_{\upgamma}(p)$ of Eq.~\eqref{eq:dispersion-kappa} occur in the CPT-even part of the electromagnetic sector of the SME, which is parametrized by $k_F$; see Eq.~\eqref{eq:photon-sector-sme}. Vacuum birefringence is suppressed at first order for a $k_F$ expressed as~\cite{Bailey:2004na,Altschul:2006zz}
\begin{equation}
\label{eq:nonbirefringent-ansatz}
(k_F)^{\mu\nu\varrho\sigma}=\frac{1}{2}(\eta^{\mu\varrho}\tilde{\kappa}^{\nu\sigma}-\eta^{\mu\sigma}\tilde{\kappa}^{\nu\varrho}-\eta^{\nu\varrho}\tilde{\kappa}^{\mu\sigma}+\eta^{\nu\sigma}\tilde{\kappa}^{\mu\varrho})\,,
\end{equation}
with a symmetric and traceless $(4\times 4)$ matrix $\tilde{\kappa}$. There are configurations of $\tilde{\kappa}$ whose dispersion equation is the square of a quadratic polynomial such that Eq.~\eqref{eq:dispersion-kappa} comes into fruition. Examples are the isotropic case with
\begin{equation}
(\tilde{\kappa}^{\mu\nu})=\frac{\tilde{\kappa}_{\mathrm{tr}}}{2}\mathrm{diag}(3,1,1,1)\,,
\end{equation}
and the nonbirefringent, anisotropic sector described by
\begin{equation}
(\tilde{\kappa}^{\mu\nu})=\frac{\tilde{\kappa}_3}{2}\mathrm{diag}(1,-1,-1,3)\,,
\end{equation}
with scalar coefficients $\tilde{\kappa}_{\mathrm{tr}}$ and $\tilde{\kappa}_3$~\cite{Kaufhold:2007qd,Klinkhamer:2010zs,Klinkhamer:2011ez}. The inverse effective metrics are diagonal and have the form
\begin{equation}
(\Omega^{\mu\nu})=\mathrm{diag}\Big[1+\tilde{\kappa}_{\mathrm{tr}},-(1-\tilde{\kappa}_{\mathrm{tr}}),-(1-\tilde{\kappa}_{\mathrm{tr}}),-(1-\tilde{\kappa}_{\mathrm{tr}})\Big]\,,
\end{equation}
and
\begin{equation}
(\Omega^{\mu\nu})=\mathrm{diag}\Big[1+\tilde{\kappa}_3,-(1+\tilde{\kappa}_3),-(1+\tilde{\kappa}_3),-(1-\tilde{\kappa}_3)\Big]\,,
\end{equation}
respectively. The effective metrics $\Omega_{\mu\nu}$ in position space straightforwardly follow from the latter.

\subsection{Nondegenerate quadratic dispersions}

Birefringence is an effect in photons that is analogous to spin nondegeneracy for fermions. There are SME background field configurations that imply two distinct photon dispersion relations, i.e., two distinct effective metrics in energy-momentum space. As a consequence, there are two effective metrics $\Omega_{\mu\nu}^{(1,2)}$ in position space, too. In principle, such a setting can be called bimetric. The two nullcones in position space are written as
\begin{equation}
\label{eq:modified-nullcones-double}
0=\Omega_{\mu\nu}^{(1,2)}\dot{x}^{\mu}\dot{x}^{\nu}\,.
\end{equation}
Thus, each polarization degree of freedom at the classical point-particle level is described by a Lagrangian of the form
\begin{equation}
\label{eq:lagrangians-kappa-12}
\tilde{L}_{\upgamma}^{(1,2)}=-\frac{1}{2\mathfrak{e}}\Omega_{\mu\nu}^{(1,2)}\dot{x}^{\mu}\dot{x}^{\nu}\,.
\end{equation}
Each Lagrangian has its own canonical Hamiltonian, as expected:
\begin{subequations}
\begin{equation}
\tilde{H}^{(1,2)}_{\upgamma}=-\frac{\mathfrak{e}}{2}\mathcal{D}_{\upgamma}^{(1,2)}(p)\,,
\end{equation}
with
\begin{equation}
\label{eq:dispersion-kappa-12}
\mathcal{D}_{\upgamma}^{(1,2)}(p)=(\Omega^{(1,2)})^{\varrho\sigma}p_{\varrho}p_{\sigma}\,.
\end{equation}
\end{subequations}
This setting covers various sets of coefficients. Birefringent dispersion equations are quartic in $p_0$. For some configurations, the quartic factorizes into two quadratic dispersion equations, i.e., $\mathcal{D}_{\upgamma}^{(1)}(p)\mathcal{D}_{\upgamma}^{(2)}(p)=0$ according to Eq.~\eqref{eq:dispersion-kappa-12}.

Let us again look at Weyl fermions first. For example, the quartic dispersion equation subject to $d_{\mu\nu}$ then factorizes into two distinct quadratic dispersion equations as follows~\cite{Kostelecky:2000mm}:
\begin{align}
0&=(p^2+2d^{(\mu\nu)}p_{\mu}p_{\nu}+d^{\mu\varrho}d_{\varrho}^{\phantom{\varrho}\nu}p_{\mu}p_{\nu}) \notag \\
&\phantom{{}={}}\times (p^2-2d^{(\mu\nu)}p_{\mu}p_{\nu}+d^{\mu\varrho}d_{\varrho}^{\phantom{\varrho}\nu}p_{\mu}p_{\nu})\,,
\end{align}
with the totally symmetric part $d^{(\mu\nu)}$ of $d^{\mu\nu}$. Hence, there are two modified Weyl cones in momentum space, which are governed by the following effective metrics:
\begin{equation}
(\Omega^{(1,2)})^{\mu\nu}=\eta^{\mu\nu}\pm 2d^{(\mu\nu)}+d^{\mu\varrho}d_{\varrho}^{\phantom{\varrho}\nu}\,.
\end{equation}
They differ at first order in Lorentz violation, but correspond to each other at second order. The effective metrics describing the modified nullcones in position space at first order in $d_{\mu\nu}$ can then be put on paper as follows:
\begin{equation}
(\Omega^{(1,2)})_{\mu\nu}=\eta_{\mu\nu}\mp 2d_{(\mu\nu)}+\dots\,.
\end{equation}
Note that for a nonzero fermion mass, the treatment of the $d_{\mu\nu}$ coefficients is highly challenging in general.

There are also birefringent $k_F$ configurations whose quartic dispersion equation factorizes. Note that Eq.~\eqref{eq:nonbirefringent-ansatz} parametrizes nonbirefringent sectors at first order in the controlling coefficients, but birefringence is possible at second order. The following parametrization for $\tilde{\kappa}$ in terms of two four-vectors $v^{\mu}$ and $w^{\mu}$ covers a wide range of such cases:
\begin{equation}
\label{eq:parameterization-2-vectors}
\tilde{\kappa}^{\mu\nu}=\frac{1}{2}(v^{\mu}w^{\nu}+v^{\nu}w^{\mu})-\frac{1}{4}(v\cdot w)\eta^{\mu\nu}\,.
\end{equation}
The two effective metrics are then~\cite{Casana:2010nd}
\begin{subequations}
\label{eq:inverse-effective-metrics-uv}
\begin{align}
(\Omega^{(1)})^{\mu\nu}&=\left(1-\frac{v\cdot w}{2}\right)\eta^{\mu\nu}+\frac{1}{2}(v^{\mu}w^{\nu}+w^{\mu}v^{\nu})\,, \\[1ex]
(\Omega^{(2)})^{\mu\nu}&=\left(1-\frac{v^2w^2}{4}\right)\eta^{\mu\nu}+\frac{1}{2}(v^{\mu}w^{\nu}+w^{\mu}v^{\nu}) \notag \\
&\phantom{{}={}}+\frac{1}{4}(v^2w^{\mu}w^{\nu}+w^2v^{\mu}v^{\nu})\,.
\end{align}
\end{subequations}
The \textit{ansatz}~\eqref{eq:parameterization-2-vectors} includes the polarization-degenerate cases of Sec.~\ref{sec:single-nullcone} by appropriate choices of $v^{\mu}$ and $w^{\mu}$. In these cases, $(\Omega^{(1)})^{\mu\nu}=\omega (\Omega^{(2)})^{\mu\nu}$ with a constant $\omega$. However, there are also sectors where $(\Omega^{(1)})^{\mu\nu}$ differs from $(\Omega^{(2)})^{\mu\nu}$ at second order in the SME coefficients. One example, known as the parity-odd sector, is given by~\cite{Schreck:2011ai}
\begin{equation}
(v^{\mu})=\begin{pmatrix}
1 \\
0 \\
0 \\
0 \\
\end{pmatrix}\,,\quad (w^{\mu})=\begin{pmatrix}
0 \\
\tilde{\kappa}^{01} \\
\tilde{\kappa}^{02} \\
\tilde{\kappa}^{03} \\
\end{pmatrix}\,.
\end{equation}
To obtain the Lagrangians of Eq.~\eqref{eq:lagrangians-kappa-12}, the effective metrics in position space must be computed from Eq.~\eqref{eq:inverse-effective-metrics-uv}. This is a straightforward task once the corresponding $(4\times 4)$ matrix representations are available after inserting the explicit forms of $u^{\mu}$ and $v^{\mu}$ into Eq.~\eqref{eq:inverse-effective-metrics-uv}.

Note that the authors of Ref.~\cite{Pfeifer:2011tk} study a similar problem in their Sec.~V.B. They propose a pseudo-Finsler structure that is, in principle, the product of the two quadratic Lagrangians of Eq.~\eqref{eq:lagrangians-kappa-12}, modulo the prefactor. The proposed form is permissible according to their definition of a Lagrangian, which includes positive homogeneity of an arbitrary degree $\geq 2$. In contrast, we introduce two independent Lagrangians, where each is positively homogeneous of first degree when taking the \textit{einbein} into consideration. Which approach is better suited remains to be seen.

\subsection{Quartic dispersions}

The generic dispersion equation for the $k_F$ background field is a very involved quartic. It can be expressed in the form \cite{Obukhov:2000nw,Kostelecky:2009zp}
\begin{subequations}
\label{eq:quartic-photon-dispersion-equation}
\begin{align}
0&=\mathcal{G}^{\mu\nu\varrho\sigma}p_{\mu}p_{\nu}p_{\varrho}p_{\sigma}\,, \\[1ex]
\mathcal{G}^{\mu\nu\varrho\sigma}&=\frac{1}{4!}\varepsilon_{\alpha\beta\gamma\delta}\varepsilon_{\zeta\eta\kappa\lambda}\chi^{\alpha\beta\zeta(\mu}\chi^{\nu|\gamma\eta|\varrho}\chi^{\sigma)\delta\kappa\lambda}\,, \\[1ex]
\chi^{\mu\nu\varrho\sigma}&=\frac{1}{2}(\eta^{\mu\varrho}\eta^{\nu\sigma}-\eta^{\nu\varrho}\eta^{\mu\sigma})+(k_F)^{\mu\nu\varrho\sigma}\,.
\end{align}
\end{subequations}
Here, $\mathcal{G}^{\mu\nu\varrho\sigma}$ is known as the Tamm-Rubilar tensor, which involves the Levi-Civita symbol $\varepsilon^{\mu\nu\varrho\sigma}$ in Minkowski spacetime. Symmetrization is performed over indices enclosed by parentheses, where excluded indices are indicated by vertical lines.

For various reasons, it has proven challenging to derive type-2 Lagrangians for quartic photon dispersion equations that do not factorize. Equation~\eqref{eq:lagrangian-standard-form-2} insinuates that the standard quadratic form is to be replaced by a corresponding quartic in position space. A suitable type-2 Lagrangian that is invariant under reparametrizations of the particle trajectory (see App.~\ref{app:reparametrization-invariance}) could take the form
\begin{equation}
\label{eq:lagrangian-type2-quartic}
\tilde{L}_{\upgamma}=-\frac{1}{2\mathfrak{e}^3}\Xi_{\mu\nu\varrho\sigma}\dot{x}^{\mu}\dot{x}^{\nu}\dot{x}^{\varrho}\dot{x}^{\sigma}\,.
\end{equation}
The totally symmetric four-tensor $\Xi_{\mu\nu\varrho\sigma}$ plays the role of an effective ``supermetric,'' i.e., a generalization of the effective metrics in modified nullcones in position space; see the previous Eqs.~\eqref{eq:modified-nullcone-single}, \eqref{eq:modified-nullcones-double}.

While effective metrics in position and momentum space are inverses of each other, the relationship between the Tamm-Rubilar tensor of the quartic dispersion equation~\eqref{eq:quartic-photon-dispersion-equation} and the effective ``supermetric'' of Eq.~\eqref{eq:lagrangian-type2-quartic} is challenging to establish. This is even true for explicit examples. To derive Eq.~\eqref{eq:lagrangian-type2-quartic} from the dispersion equation, as we did for the quadratic cases, the canonical momentum must be inverted for the four-velocity. This amounts to solving a system of cubic equations, which is highly challenging.

Interestingly, we can follow an alternative path that benefits from the Fresnel ray equation of material optics. By employing the latter and basic arguments on classical-particle propagation in position space, we are able to derive a classical type-2 Lagrangian in the form of Eq.~\eqref{eq:lagrangian-type2-quartic} for a vacuum with specific properties; see App.~\ref{app:quartic-em-dispersion}.

Note that Skakala and Visser worked on pseudo-Finsler structures associated with light propagation in biaxial materials~\cite{Skakala:2008kf,Skakala:2008jp}. Optical properties of the latter are encoded in a manifestly quartic dispersion equation according to Eq.~\eqref{eq:quartic-photon-dispersion-equation} that does not factorize into quadratic ones. They encountered the severe problem of singular pseudo-Finsler metrics, not only along the optical axes but for generic lightlike four-velocities and momenta, respectively. Their conclusion is discouraging and, in principle, states that biaxial materials did not prove to be the testing ground for quantum-gravity analogs that they had been hoping for.

Lagrangians of the form of Eq.~\eqref{eq:lagrangian-type2-quartic} might turn out to be more suitable for this endeavor. However, because of the challenges that we described above, the development of a generic approach to classical point-particle Lagrangians for quartic photon dispersion equations, as well as embedding them into a well-defined pseudo-Finslerian setting, still presents an interesting open problem.

\subsection{Carroll-Field-Jackiw theory}

The complementary CPT-odd photon sector of the SME is governed by a vector-valued background field $k_{AF}^{\mu}$; see Eq.~\eqref{eq:photon-sector-sme}~\cite{Carroll:1989vb,Colladay:1998fq,Adam:2001ma}. In the literature, it is known as either Carroll-Field-Jackiw (CFJ) theory, in homage to the physicists who studied it first, or as Maxwell-Chern-Simons theory. The latter name is due to the form of the Lagrange density, which resembles that of Chern-Simons theory in $(2+1)$ spacetime dimensions~\cite{Chern:1974}. However, contrary to the $k_{AF}^{\mu}$ sector of the SME, a genuine Chern-Simons term does not violate Lorentz invariance and is a topological field theory.

The dispersion equation is quartic in the momentum and resembles that of the $b_{\mu}$ coefficients stated in Eq.~\eqref{eq:quartic-dispersion-b}:
\begin{subequations}
\begin{align}
\mathcal{D}_{\mathrm{CFJ}}(p)&=0\,, \\[1ex]
\mathcal{D}_{\mathrm{CFJ}}(p)&=p^4-[(k_{AF}\cdot p)^2-k_{AF}^2p^2]\,,
\end{align}
\end{subequations}
where $k_{AF}^{\mu}$ has mass dimension 1. In analogy to how we treated the $b_{\mu}$ coefficients, the latter dispersion equation can be decomposed as follows:
\begin{subequations}
\begin{align}
\mathcal{D}_{\mathrm{CFJ}}(p)&=\mathcal{D}_{\mathrm{CFJ}}^{(+)}(p)\mathcal{D}_{\mathrm{CFJ}}^{(-)}(p)\,, \\[1ex]
\mathcal{D}_{\mathrm{CFJ}}^{(\pm)}(p)&=p^2\pm\sqrt{(k_{AF}\cdot p)^2-k_{AF}^2p^2}\,.
\end{align}
\end{subequations}
A comparison to Eq.~\eqref{eq:dispersions-b} reveals the correspondences
\begin{equation}
b^{\mu}\leftrightarrow\frac{k_{AF}^{\mu}}{2}\,,\quad m^2\leftrightarrow -\frac{k_{AF}^2}{4}\,,
\end{equation}
between the background fields and the fermion mass. This allows us to infer that the type-2 Lagrangians of the $k_{AF}^{\mu}$ sector must be of the form
\begin{equation}
\label{eq:lagrangian-second-form-kAF}
\tilde{L}_{\mathrm{CFJ}}^{(\pm)}=-\frac{1}{2}\left(\frac{\dot{x}^2}{\mathfrak{e}}\pm \sqrt{(k_{AF}\cdot \dot{x})^2-k_{AF}^2\dot{x}^2}-\frac{\mathfrak{e}}{4}k_{AF}^2\right)\,.
\end{equation}
Note the presence of $k_{AF}^2/4$, which mimics a fermion mass, albeit with wrong sign. The latter classical Lagrangians are manifestly massless, though.

\section{Conclusions and outlook}
\label{eq:conclusions}

This paper has been written as an extension of contemporary literature on classical, relativistic, point-particle Lagrangians based on the SME. Our results rest upon a Lagrangian of a form not widely considered in the SME community. The derived Lagrangians have a significant advantage over the known ones in the literature: they possess a well-defined massless limit. Hence, they are suitable for studying classical-particle analogs to the SME photon sector.

We found alternative Lagrangians for the spin-degenerate $a_{\mu}$ and $e_{\mu}$ coefficients, as well as the spin-nondegenerate $b_{\mu}$ and $H_{\mu\nu}$ coefficients. Further results for massless fermions were presented for the $c_{\mu\nu}$ and $d_{\mu\nu}$ coefficients. These novel Lagrangians possess a different constraint structure compared to those known in the literature, which is analyzed in a dedicated study. As expected, the number of physical degrees of freedom of a pointlike particle amounts to 3, even in the presence of Lorentz violation. Contrary to the type-1 Lagrangians published previously, the canonical Hamiltonians are \emph{not} identically equal to 0, but they are directly related to the dispersion equations of the settings studied.

A particular interest of ours was in classical-particle analogs of photons. These have been challenging to construct according to the procedure used over the past 15 years, which strongly relies on particles being massive. For degenerate quadratic dispersion equations or quartic equations factorizing into quadratic ones, the Lagrangians are straightforward extensions of the standard one. The latter involve the modified nullcone(s) in position space, which can be derived from the quadratic dispersion equations in a direct manner. To obtain Lagrangians for a quartic dispersion equation has proven to be challenging. A generic and suitable method of treating those must still be developed.

The Lagrangians obtained here may be applied to classical-particle analogs of Weyl fermions and photons, subject to spacetime symmetry violation and propagating in gravitational fields. In particular, we refer to problems of massless-particle propagation in black-hole and wormhole solutions of modified-gravity theories with spacetime symmetry violation.

Another intriguing direction worth pursuing is the relationship between the classical-Lagrangians found and (pseudo-)Finsler geometry. The latter is an extension of (pseudo)-Riemannian geometry that supports path length intervals with intrinsic vector and tensor fields. Over the years, the relationship between type-1 Lagrangians within the SME and Finsler geometry has been pointed out in various papers. In fact, classical particles subject to Lorentz violation, as described by the SME, propagate along geodesics in pseudo-Finsler spaces~\cite{Kostelecky:2011qz}. Extending these and similar analyses to the type-2 Lagrangians introduced here presents another interesting research project.

\section*{Acknowledgments}

MS is grateful to the hospitality and support by the Department of Exact and Natural Sciences of the State University of Southwest Bahia (UESB), where the initial spark of this work ignited. MS is also indebted to CNPq Produtividade 307653/2025-0 and CAPES/Finance Code 001. JAASR and RT express their gratitude for the partial financial assistance provided by the UESB through Grant AuxPPI Edital No. 267/2024. JAASR is also indebted to FAPESB-CNPq/Produtividade 12243/2025 (TOB-BOL2798/2025).

\appendix

\section{Reparametrization invariance}
\label{app:reparametrization-invariance}

Here, we intend to demonstrate the reparametrization invariance of two exemplary type-2 Lagrangians. First, let us look at a generic modified type-2 Lagrangian that involves an effective metric $\Omega_{\mu\nu}$ and is of the form
\begin{equation}
\tilde{L}_1=\frac{1}{2}\left(\frac{\Omega_{\mu\nu}\dot{x}^{\mu}\dot{x}^{\nu}}{\mathfrak{e}}+m^2\mathfrak{e}\right)\,.
\end{equation}
Under reparametrizations of the particle trajectory described by a function $f=f(\lambda)$, we use the variational properties $\delta x^{\mu}=f\dot{x}^{\mu}$ and $\delta\mathfrak{e}=f\dot{\mathfrak{e}}+\dot{f}\mathfrak{e}$~\cite{Brink:1976uf,Tong:2009np}. Then, the variation of the Lagrangian is given by
\begin{align}
\delta\tilde{L}_1&=\frac{1}{2}\left(\frac{2\Omega_{\mu\nu}\dot{x}^{\mu}}{\mathfrak{e}}\frac{\mathrm{d}(\delta x^{\nu})}{\mathrm{d}\lambda}-\frac{\Omega_{\mu\nu}\dot{x}^{\mu}\dot{x}^{\nu}}{\mathfrak{e}^2}\delta \mathfrak{e}+m^2\delta \mathfrak{e}\right) \notag \\
&=\frac{1}{2}\bigg[\frac{2\Omega_{\mu\nu}\dot{x}^{\mu}}{\mathfrak{e}}\frac{\mathrm{d}(f\dot{x}^{\nu})}{\mathrm{d}\lambda}-\frac{\Omega_{\mu\nu}\dot{x}^{\mu}\dot{x}^{\nu}}{\mathfrak{e}^2}(f\dot{\mathfrak{e}}+\dot{f}\mathfrak{e}) \notag \\
&\phantom{{}={}}\hspace{0.5cm}+m^2(f\dot{\mathfrak{e}}+\dot{f}\mathfrak{e})\bigg] \notag \\
&=\frac{f}{2}\bigg(2\Omega_{\mu\nu}\dot{x}^{\mu}\frac{\ddot{\phi}^{\nu}}{\mathfrak{e}}-\frac{\Omega_{\mu\nu}\dot{x}^{\mu}\dot{x}^{\nu}}{\mathfrak{e}^2}\dot{\mathfrak{e}}+m^2\dot{\mathfrak{e}}\bigg) \notag \\
&\phantom{{}={}}+\frac{\dot{f}}{2}\left(\frac{2\Omega_{\mu\nu}\dot{x}^{\mu}\dot{x}^{\nu}}{\mathfrak{e}}-\frac{\Omega_{\mu\nu}\dot{x}^{\mu}\dot{x}^{\nu}}{\mathfrak{e}}+m^2\mathfrak{e}\right) \notag \\
&=f\dot{\tilde{L}}_1+\dot{f}\tilde{L}_1=\frac{\mathrm{d}}{\mathrm{d}\lambda}(f\tilde{L}_1)\,,
\end{align}
i.e., it is a total derivative.

Next, we investigate a type-2 Lagrangian governed by a ``supermetric'' $\Xi_{\mu\nu\varrho\sigma}$; cf.~Eq.~\eqref{eq:lagrangian-type2-quartic}:
\begin{equation}
\tilde{L}_2=\frac{1}{2\mathfrak{e}^3}\Xi_{\mu\nu\varrho\sigma}\dot{x}^{\mu}\dot{x}^{\nu}\dot{x}^{\varrho}\dot{x}^{\sigma}\,.
\end{equation}
Varying the latter leads to
\begin{align}
\delta\tilde{L}_2&=\frac{1}{2}\bigg[\frac{4}{\mathfrak{e}^3}\Xi_{\mu\nu\varrho\sigma}\dot{x}^{\mu}\dot{x}^{\nu}\dot{x}^{\varrho}\frac{\mathrm{d}(\delta x^{\sigma})}{\mathrm{d}\lambda}-\frac{3}{\mathfrak{e}^4}\Xi_{\mu\nu\varrho\sigma}\dot{x}^{\mu}\dot{x}^{\nu}\dot{x}^{\varrho}\dot{x}^{\sigma}\delta \mathfrak{e}\bigg] \notag \displaybreak[0]\\
&=\frac{1}{2}\bigg[\frac{4}{\mathfrak{e}^3}\Xi_{\mu\nu\varrho\sigma}\dot{x}^{\mu}\dot{x}^{\nu}\dot{x}^{\varrho}\frac{\mathrm{d}(f\dot{x}^{\sigma})}{\mathrm{d}\lambda} \notag \\
&\phantom{{}={}}\hspace{0.5cm}-\frac{3}{\mathfrak{e}^4}\Xi_{\mu\nu\varrho\sigma}\dot{x}^{\mu}\dot{x}^{\nu}\dot{x}^{\varrho}\dot{x}^{\sigma}(f\dot{\mathfrak{e}}+\dot{f}\mathfrak{e})\bigg] \notag \displaybreak[0]\\
&=\frac{f}{2}\bigg(\frac{4}{\mathfrak{e}^3}\Xi_{\mu\nu\varrho\sigma}\dot{x}^{\mu}\dot{x}^{\nu}\dot{x}^{\varrho}\ddot{x}^{\sigma}-\frac{3}{\mathfrak{e}^4}\Xi_{\mu\nu\varrho\sigma}\dot{x}^{\mu}\dot{x}^{\nu}\dot{x}^{\varrho}\dot{x}^{\sigma}\dot{\mathfrak{e}}\bigg) \notag \\
&\phantom{{}={}}+\frac{\dot{f}}{2}\left(\frac{4}{\mathfrak{e}^3}\Xi_{\mu\nu\varrho\sigma}\dot{x}^{\mu}\dot{x}^{\nu}\dot{x}^{\varrho}\dot{x}^{\sigma}-\frac{3}{\mathfrak{e}^3}\Xi_{\mu\nu\varrho\sigma}\dot{x}^{\mu}\dot{x}^{\nu}\dot{x}^{\varrho}\dot{x}^{\sigma}\right) \notag \\
&=f\dot{\tilde{L}}_2+\dot{f}\tilde{L}_2=\frac{\mathrm{d}}{\mathrm{d}\lambda}(f\tilde{L}_2)\,,
\end{align}
which also amounts to a total derivative.

\section{Additional type-2 Lagrangians}
\label{app:additional-lagrangians}

In the forthcoming sections, we will present additional examples of type-2 Lagrangians based on the Dirac fermion sector of the SME. These findings support the results in the main body of the paper. However, they do not provide anything substantially different from what was already considered, which is why we opted to separate them from the main text.

\subsection{Spin-degenerate $c_{\mu\nu}$}

The minimal-SME coefficients $c_{\mu\nu}$ provide a skewed effective metric at first order in these coefficients:
\begin{equation}
\label{eq:effective-metric}
\Omega_{\mu\nu}=\eta_{\mu\nu}+2c_{(\mu\nu)}\,,
\end{equation}
where $c_{(\mu\nu)}$ is the symmetric part of $c_{\mu\nu}$. The proposed type-2 Lagrangian is
\begin{equation}
\tilde L_c=-\frac12 \left(\frac{\Omega_{\mu\nu} \dot x^\mu \dot x^\nu}{\mathfrak e}+ \mathfrak e m^2\right)\,.
\end{equation}
The canonical momentum follows directly from differentiation:
\begin{equation}
p_\mu=-\frac{\partial \tilde L_c}{\partial \dot x^\mu}=\frac{\Omega_{\mu\nu}\dot x^\nu}{\mathfrak e}\,.
\end{equation}
The effective metric of Eq.~\eqref{eq:effective-metric} is invertible if $|c_{\mu\nu}|\ll 1$. Then, the velocity can be expressed in terms of the momentum as
\begin{equation}
\dot x^\mu=\mathfrak e\,\Omega^{\mu\nu}p_\nu\,,
\end{equation}
where $\Omega^{\mu\nu}\Omega_{\nu\varrho}=\delta^{\mu}_{\phantom{\mu}\varrho}$. The Legendre transformation results in the canonical Hamiltonian
\begin{subequations}
\begin{align}
\tilde H_c&=-p\cdot \dot{x} -\tilde L_c \notag \\
&=-\frac{1}{2}\frac{\Omega_{\mu\nu} \dot x^\mu\dot x^\nu}{\mathfrak e}+\frac{1}{2}\mathfrak e m^2=-\frac{\mathfrak e}{2}\mathcal{D}_c(p)\,,
\end{align}
where
\begin{equation}
\mathcal{D}_c(p)=\Omega^{\mu\nu} p_\mu p_\nu-m^2\,.
\end{equation}
\end{subequations}
Thus, the effect of $c_{\mu\nu}$ in both $\tilde{L}_c$ and $\tilde{H}_c$ is to act as an effective metric. In position space for $m=0$, it describes a modified nullcone along which particle propagation occurs. The inverse effective metric in momentum space also governs a modified nullcone with complementary properties. Multiple branches do not exist, since the $c$ coefficients are spin-degenerate.

\subsection{Subset of spin-nondegenerate $d_{\mu\nu}$}

We propose the following type-2 Lagrangian for the $d_{\mu\nu}$ coefficients:
\begin{equation}
\label{eq:Ld-type2}
\tilde{L}_d^{(\pm)}=-\frac{1}{2}\left(\frac{\dot{x}^2}{\mathfrak{e}}\pm 2m\sqrt{\dot{x}\cdot d^{2}\cdot\dot{x}}+\mathfrak{e}m^2\right)\,,
\end{equation}
where $(d^2)_{\mu\nu}:=d_{\mu}{}^{\alpha}d_{\alpha\nu}$ and \(\dot x\cdot d^{2}\cdot\dot x:= \dot x^\mu (d^2)_{\mu\nu}\dot x^\nu\). The canonical momentum is obtained from
\begin{equation}
p_\mu=-\frac{\partial \tilde L_d^{(\pm)}}{\partial \dot x^\mu}\,.
\end{equation}
The first term in Eq.~\eqref{eq:Ld-type2} yields
\begin{equation}
-\frac{\partial}{\partial \dot{x}^\mu}\left(-\frac{1}{2}\frac{\dot x^2}{\mathfrak e}\right)=
\frac{1}{\mathfrak e}\,\dot x_\mu\,.
\end{equation}
Let us define the short-hand notation \(S := \sqrt{\dot x\cdot d^2\cdot\dot x}\) for the square-root term, whereupon
\begin{equation}
-\frac{\partial}{\partial \dot x^\mu}(\mp m S)=\pm m\,\frac{(d^2)_{\mu\nu}\dot x^\nu}{S}\,.
\end{equation}
Compiling both contributions implies the canonical momentum
\begin{equation}
p_\mu=\frac{\dot x_\mu}{\mathfrak e}
\pm m\,\frac{(d^2)_{\mu\nu}\dot x^\nu}{\sqrt{\dot x\cdot d^2\cdot\dot x}}\,.
\label{eq:pd}
\end{equation}
This expression is invertible for $\dot x^\mu$, as in the $b_\mu$ and $H_{\mu\nu}$ cases. Defining the Gramian
\(R := p\cdot d^2 \cdot p\), we can invert Eq.~\eqref{eq:pd} as we did for $b_{\mu}$:
\begin{equation}
\dot x^\mu=\mathfrak e\left(p^\mu \mp m\,\frac{(d^2)^{\mu\nu}p_{\nu}}{\sqrt{p\cdot d^2\cdot p}}\right)\,.
\label{eq:vel-d}
\end{equation}
The latter can be checked by inserting it into Eq.~\eqref{eq:pd}. The canonical Hamiltonians then follow from the Legendre transformation
\begin{equation}
\tilde H_d^{(\pm)}= - p\cdot\dot x - \tilde L_d^{(\pm)}\,.
\end{equation}
Inserting Eq.~\eqref{eq:vel-d} into the first term gives rise to
\begin{align}
- p\cdot\dot x&=-\mathfrak e\left(p^2\mp m\frac{p\cdot d^{2}\cdot p}{\sqrt{p\cdot d^{2}\cdot p}}\right)
\notag \\
&=-\mathfrak e p^2 \pm \mathfrak e m\sqrt{p\cdot d^{2}\cdot p}\,.
\end{align}
Moreover, from Eq.~\eqref{eq:vel-d}, we arrive at
\begin{equation}
\frac{\dot x^2}{\mathfrak e}
= \mathfrak e\,p^2\,,\quad
\sqrt{\dot x\cdot d^{2}\cdot\dot x}
= \mathfrak e\,\sqrt{p\cdot d^{2}\cdot p}\,,
\end{equation}
which allows us to formulate the canonical Hamiltonians:
\begin{subequations}
\begin{align}
\tilde H_d^{(\pm)}
&=-\left(-\mathfrak e p^2 \pm \mathfrak e m\sqrt{p\cdot d^{2}\cdot p}\right)
\notag \\
&\phantom{{}={}}+\frac12\left(\mathfrak e p^2\pm 2\mathfrak e m\sqrt{p\cdot d^{2}\cdot p}+\mathfrak e m^2\right) \notag \\
&=-\frac{\mathfrak e}{2}\mathcal{D}_d^{(\mp)}(p)\,,
\end{align}
with
\begin{equation}
\mathcal{D}_d^{(\pm)}(p)=p^2 - m^2\pm 2m\sqrt{p\cdot d^{2}\cdot p}\,.
\end{equation}
\end{subequations}
Notably, the two Hamiltonians differ by the sign before the square root of the Gramian. Multiplying both branches yields the complete quartic dispersion equation:
\begin{equation}
\mathcal{D}_d^{(+)}\mathcal{D}_d^{(-)}=(p^2-m^2)^2-4m^2\,p\cdot d^{2}\cdot p\,,
\end{equation}
which is, in fact, equivalent to the full SME dispersion law for the $d_{\mu\nu}$ coefficients.

\subsection{Combined set $H_{\mu\nu}+d_{\mu\nu}$}

It is even possible to derive Lagrangians for certain combinations of coefficients. For a Lorentz-violating setting that involves both the antisymmetric tensor $H_{\mu\nu}$ and the symmetric tensor $d_{\mu\nu}$, the relevant
invariant appearing in the square-root structure is
\begin{equation}
\Delta_{Hd}(\dot x)=\dot{x}\cdot H\cdot H\cdot \dot{x}+ 2X \dot{x}^2+ m^2\, \dot{x}\cdot d^{2}\cdot \dot{x}\,,
\end{equation}
with $X$ given in Eq.~\eqref{definition-X-Y} and $(d^{2})_{\mu\nu}:=d_{\mu\alpha}d^{\alpha}{}_{\nu}$. The corresponding type-2 Lagrangians are
\begin{equation}
\tilde L_{Hd}^{(\pm)}=-\frac12\left(\frac{\dot x^2}{\mathfrak e}\pm2\sqrt{\Delta_{Hd}(\dot x)}+ \mathfrak e m^2\right)\,,
\end{equation}
which reduce continuously to the pure-$H$ and pure-$d$ Lagrangians of Eqs.~\eqref{eq:lagrangian-second-form-H} and \eqref{eq:Ld-type2}, respectively, when the complementary coefficients are discarded.

Differentiation yields the canonical momentum
\begin{align}
p_\mu=&-\frac{\partial \tilde L_{Hd}^{(\pm)}}{\partial \dot x^\mu} \notag \\
&=\frac{\dot x_\mu}{\mathfrak e}\pm\frac{(HH)_{\mu\nu}\dot x^\nu+2X\dot x_\mu+m^2(d^{2})_{\mu\nu}\dot x^\nu}{
\sqrt{\dot{x}\cdot H\cdot H\cdot \dot{x}+2X\dot{x}^2+ m^{2}\,\dot{x}\cdot d^{2}\cdot \dot{x}}} \notag \\
&=\frac{\dot x_\mu}{\mathfrak e}
\pm \frac{K_{\mu\nu}\dot x^\nu}{\sqrt{\Delta_{Hd}(\dot x)}}\,,
\end{align}
which we expressed in terms of the tensor
\begin{equation}
K_{\mu\nu}:=(HH)_{\mu\nu}+2X\eta_{\mu\nu}+ m^2 (d^{2})_{\mu\nu}\,.
\end{equation}
The standard SME Gramian identity allows for inverting $p_{\mu}$ for the particle velocity:
\begin{equation}
\dot x^\mu=\mathfrak e\left(p^\mu\mp\frac{K^{\mu\nu} p_{\nu}}{\sqrt{p\cdot K\cdot p}}\right)\,.
\end{equation}
For the Legendre transformation, the following findings are valuable:
\begin{subequations}
\begin{align}
p_\mu \dot x^\mu&=\mathfrak e\left(p^2\mp \sqrt{p\cdot K\cdot p}\right)\,, \\[1ex]
\sqrt{\Delta_{Hd}(\dot x)}&=\mathfrak e\sqrt{p\cdot K\cdot p}\,.
\end{align}
\end{subequations}
Then, the resulting canonical Hamiltonians read
\begin{equation}
\tilde H_{Hd}^{(\pm)}=\frac{\mathfrak e}{2}\left(-p^2 + m^2\pm 2\sqrt{p\cdot K\cdot p}\,\right)\,.
\end{equation}
These Hamiltonians exhibit two branches, interpolating smoothly between the pure-$H$ and pure-$d$ cases. Their product reproduces the appropriate minimal-sector limit of the general SME dispersion equation given in Eq.~(40) of Ref.~\cite{Kostelecky:2013rta}.

\section{Quartic electromagnetic dispersion}
\label{app:quartic-em-dispersion}

Let us consider a light ray propagating along a quartic surface expressed as
\begin{subequations}
\label{eq:quartic-surface}
\begin{align}
0&=f(t,x,y,z)\,, \\[1ex]
f(t,x,y,z)&=\Xi_{\mu\nu\varrho\sigma}x^{\mu}x^{\nu}x^{\varrho}x^{\sigma}\,.
\end{align}
\end{subequations}
This scenario describes a nontrivial biaxial vacuum. The surface consists of two sheets that intersect each other along the optical axes. Hence, it is singular along these directions, which are excluded from the analysis.

The trajectory $x^{\mu}=x^{\mu}(\lambda)$ of a classical point-particle analog runs along the surface, which is why it must satisfy the previous equation. Since the particle moves without external force, $\frac{\mathrm{d}^nx^{\mu}}{\mathrm{d}\lambda^n}=0$ for $n\geq 2$ and $x^{\mu}$ is collinear with $\dot{x}^{\mu}$. Thus, differentiating Eq.~\eqref{eq:quartic-surface} four times for $\lambda$ implies that the four-velocity $\dot{x}^{\mu}$ also obeys Eq.~\eqref{eq:quartic-surface}:
\begin{align}
\label{eq:quartic-surface-four-velocity}
0&=f(\dot{t},\dot{x},\dot{y},\dot{z})\,, \\[1ex]
f(\dot{t},\dot{x},\dot{y},\dot{z})&=\Xi_{\mu\nu\varrho\sigma}\dot{x}^{\mu}\dot{x}^{\nu}\dot{x}^{\varrho}\dot{x}^{\sigma}\,.
\end{align}
The relationship between the quartic SME dispersion equation and Eq.~\eqref{eq:quartic-surface-four-velocity} in position space is not directly evident. However, we can borrow a result from material optics, which is known as the Fresnel ray equation~\cite{Knoerrer:1986}:
\begin{equation}
0=\frac{x^2}{1/v_1^2-1/v^2}+\frac{y^2}{1/v_2^2-1/v^2}+\frac{z^2}{1/v_3^2-1/v^2}\,.
\end{equation}
Here, $v$ is the norm of the velocity of a light ray at the point $\mathbf{x}=(x,y,z)$ in the coordinate system of principal axes. The velocity component along the $i$-th principal axis is denoted as $v_i$. Substituting $v=|\mathbf{x}|/t$ leads to the following equivalent quartic equation in $t$ and $\mathbf{x}$:
\begin{subequations}
\begin{align}
0&=v_1^2v_2^2v_3^2t^4-t^2\tilde{\beta}(\mathbf{x})+\tilde{\gamma}(\mathbf{x})\,, \\[1ex]
\tilde{\beta}(\mathbf{x})&=v_1^2(v_2^2+v_3^2)x^2+v_2^2(v_1^2+v_3^2)y^2 \notag \\
&\phantom{{}={}}+v_3^2(v_1^2+v_2^2)z^2\,, \\[1ex]
\tilde{\gamma}(\mathbf{x})&=\mathbf{x}^2(v_1^2x^2+v_2^2y^2+v_3^2z^2)\,.
\end{align}
\end{subequations}
We consider a Lorentz-violating vacuum characterized by the property $v_i=1/\varepsilon_i$, where $\varepsilon_i$ is the $i$-th component of the permittivity tensor in the principal coordinate system. Expressing the velocity components in terms of the permittivity allows us to reformulate the latter quartic equation once more:
\begin{subequations}
\begin{align}
0&=t^4-t^2\beta(\mathbf{x})+\gamma(\mathbf{x})\,, \\[1ex]
\beta(\mathbf{x})&=(\varepsilon_2+\varepsilon_3)x^2+(\varepsilon_1+\varepsilon_3)y^2+(\varepsilon_1+\varepsilon_2)z^2\,, \\[1ex]
\gamma(\mathbf{x})&=\mathbf{x}^2(\varepsilon_2\varepsilon_3x^2+\varepsilon_1\varepsilon_3y^2+\varepsilon_1\varepsilon_2z^2)\,.
\end{align}
\end{subequations}
According to the previous arguments, the four-velocity~$\dot{x}^{\mu}$ of the classical-particle analog must satisfy the very same quartic equation. Then, we can propose a classical point-particle Lagrangian in the form of Eq.~\eqref{eq:lagrangian-type2-quartic} with the nonzero independent components of the totally symmetric ``supermetric''
\begin{subequations}
\begin{align}
\Xi_{0000}&=1\,,\quad \Xi_{0011}=-\frac{1}{6}(\varepsilon_2+\varepsilon_3)\,, \displaybreak[0]\\[1ex]
\Xi_{0022}&=-\frac{1}{6}(\varepsilon_1+\varepsilon_3)\,,\quad \Xi_{0033}=-\frac{1}{6}(\varepsilon_1+\varepsilon_2)\,, \displaybreak[0]\\[1ex]
\Xi_{1111}&=\varepsilon_2\varepsilon_3\,,\quad \Xi_{2222}=\varepsilon_1\varepsilon_3\,,\quad \Xi_{3333}=\varepsilon_1\varepsilon_2\,, \displaybreak[0]\\[1ex]
\Xi_{1122}&=\frac{1}{6}(\varepsilon_1+\varepsilon_2)\varepsilon_3\,,\quad \Xi_{1133}=\frac{1}{6}(\varepsilon_1+\varepsilon_3)\varepsilon_2\,, \displaybreak[0]\\[1ex]
\Xi_{2233}&=\frac{1}{6}(\varepsilon_2+\varepsilon_3)\varepsilon_1\,,
\end{align}
\end{subequations}
where the remaining ones vanish.


\end{document}